\documentclass[aps,nofootinbib]{revtex4}

\usepackage{graphics}
\usepackage{amsmath}
\usepackage{amsfonts}

\begin{document}

\title{A renormalisation group approach to two-body scattering in the presence
of long-range forces}

\author{Thomas Barford}
\author{Michael C. Birse}

\affiliation{Theoretical Physics Group, Department of Physics and Astronomy\\
University of Manchester, Manchester, M13 9PL, UK\\}
\date{\today}


\begin{abstract}
We apply renormalisation-group methods to two-body scattering by a combination 
of known long-range and unknown short-range potentials. We impose a cut-off 
in the basis of distorted waves of the long-range potential and identify 
possible fixed points of the short-range potential as this cut-off is lowered 
to zero. The expansions around these fixed points define the power countings 
for the corresponding effective field theories. Expansions around nontrivial
fixed points are shown to correspond to distorted-wave versions of the 
effective-range expansion. These methods are applied to scattering in the 
presence of Coulomb, Yukawa and repulsive inverse-square potentials.
\end{abstract}

\maketitle

\section{Introduction}

Effective field theories (EFT's) offer the promise of a systematic and 
model-independent treatment of nuclear and hadronic physics at 
low energies. In the form of chiral perturbation theory (ChPT), they
have been used with some success in both the mesonic and single-nucleon 
sectors (for reviews see: \cite{eck,bkm95}). Currently there is much
interest in extending these applications to few-nucleon systems, as reviewed
in Refs.~\cite{vk4,eft1,eft2,border}.

These theories rely on the existence of a separation of scales between those 
of interest for low-energy physics, such as momenta, energies or the pion mass,
and those of the underlying short-distance physics, such the $\rho$-meson mass,
the nucleon mass or $4\pi f_\pi$. This makes it possible to expand any 
observable systematically in powers of the ratio $Q/\Lambda_0$, where $Q$ 
denotes a generic low-energy scale and $\Lambda_0$ a typical scale of the 
underlying physics. In addition, the effective Lagrangian or Hamiltonian 
used to calculate these observables can be organised in a similar way. 
Such an expansion will be useful provided that the separation of scales is wide
enough for the expansion to converge rapidly.

An EFT is defined by a Lagrangian or Hamiltonian containing all possible local 
terms consistent with the symmetries of the underlying theory. Although there
is an infinite number of these terms, they can be organised according to a 
``power counting", which is related to the number of low-energy scales present 
in each term. Loop diagrams are of the same or higher order compared with the 
terms from which they are constructed. This makes systematic calculations possible 
since, if we want to calculate observables up to some order in $Q/\Lambda_0$, we 
need only terms in the effective theory up to a corresponding order. 
For weakly interacting systems the terms in the expansion can be organised 
according to naive dimensional analysis, a term proportional to 
$(Q/\Lambda_0)^d$ being counted as of order $d$ \cite{wein1,wein2}. This 
``Weinberg" power counting is the one familiar from ChPT in the zero- and 
one-nucleon sectors.

In contrast, for strongly interacting systems there can be new low-energy 
scales which are generated by nonperturbative dynamics. An important example
for nuclear physics is the very large $s$-wave scattering length in 
nucleon-nucleon scattering. In such cases we need to resum certain terms in
the theory to all orders, and this leads to a new power 
counting \cite{bvk1,vk3,ksw2,geg,bmr}, often referred to as 
Kaplan, Savage and Wise (KSW) power counting.

The theoretical tool that allows us to determine the power counting in these
cases is the renormalisation group (RG), which is used to study the scaling 
behaviour of systems in a wide range of areas of physics. It is an extension
of the simple dimensional analysis which leads to Weinberg power counting
is weakly-interacting systems. 

In our work we use a Wilsonian version of the RG \cite{wrg}, imposing a 
momentum cut-off, $|{\bf k}|<\Lambda$, on the low-energy EFT. This cut-off 
should be thought of as separating the low-energy physics, which we wish to 
treat explicitly, from the underlying high-energy physics. 

The cut-off $\Lambda$ should be chosen to be above all of the low-energy scales of 
interest, but well below the scales of the underlying physics. This is possible
so long as there is a clear separation between these scales. Beyond this, the 
precise value of $\Lambda$ is arbitrary and we require that physical 
observables be independent of $\Lambda$. This means that physics on momentum 
scales higher than $\Lambda$ has not just been discarded. Rather, it has been 
``integrated out", its effects being included implicitly in the couplings of the 
EFT. As a result, all coupling constants in the EFT become functions of the cut-off 
$\Lambda$. Finally we rescale the theory, expressing all dimensioned quantities in 
units of $\Lambda$. The various coupling constants in the resulting rescaled 
theory vary with the cut-off $\Lambda$, and their flow is described by a 
first-order differential equation: the RG equation.

For a system with a clear separation of scales, the rescaled coupling constants 
become independent of $\Lambda$ as $\Lambda\rightarrow 0$. This is because, for
$\Lambda\ll\Lambda_0$, the only scale left is $\Lambda$ and so the rescaled
theory becomes independent of $\Lambda$. The theory is said to flow towards a 
fixed point of the RG.

Close to a fixed point, deviations from it scale as powers of the 
cut-off. Since the rescaling means that each low-energy scale appearing in a 
term of the potential contributes one power of $\Lambda$, we can use this to 
define the power counting for our EFT. A term that behaves like $\Lambda^\nu$ 
is assigned an order $d=\nu-1$, to match with the Weinberg power counting
mentioned above.

Perturbations around a fixed point can be classified into three types according 
to the sign of $\nu$. One with $\nu>0$ is known as an ``irrelevant" perturbation. 
Its flow is towards the fixed point as $\Lambda\rightarrow 0$. A perturbation
with $\nu=0$ is called ``marginal". This is the type of term familiar in 
conventionally renormalisable field theories. If a marginal perturbation is 
present, we should expect to find logarithmic flow with $\Lambda$. Finally, a
perturbation with $\nu<0$ is called ``relevant". It leads to flow away from the 
fixed point as $\Lambda\rightarrow 0$. 

If all perturbations about a fixed point are irrelevant, then the fixed point is 
stable: the couplings of any theory close to that point will flow towards it as 
$\Lambda\rightarrow 0$. On the other hand, if there are one or more relevant 
perturbations then the fixed point is unstable.

In Ref.~\cite{bmr} these ideas were applied to two-body scattering by short-range 
forces. Two fixed points were found for $s$-wave scattering. One is the trivial 
fixed point, describing systems with no scattering. The terms in the expansion 
around this point can be organised according to Weinberg power counting. This 
defines an EFT which is appropriate for describing a system with weak scattering.
The second point is a nontrivial one describing systems with a bound state at zero 
energy. The flow near it includes one unstable direction and the appropriate 
power counting for expanding around it is KSW counting. The terms of this 
expansion are in one-to-one correspondence with the terms of the effective-range 
expansion \cite{ere,bj,bethe,newton}.

The expansion around this nontrivial fixed point is appropriate for a system with 
a large scattering length (or equivalently a bound state or resonance close to 
zero energy). In the case of nucleon-nucleon scattering, it can be used at very
low energies, well below the pion mass, where the whole strong interaction between 
the nucleons can be treated as short-range. It has been applied successfully
to the calculation of various deuteron properties and reactions \cite{crs}. The
corresponding EFT has also been extended to describe three-body 
systems \cite{bvk1,bhvk}. (For more examples see Refs.~\cite{vk4,border}.)

At higher energies, where the nucleon momenta are comparable to the pion mass, 
one would like to treat pion-exchange forces as known long-range interactions,
calculable from ChPT \cite{orvk}. Similarly, in proton-proton scattering the Coulomb 
interaction provides a known long-range force. In both of these cases, the 
remaining shorter-range strong force between the nucleons could be described 
within an EFT in terms of contact interactions. 

Kong and Ravndal \cite{kr} have studied the example of scattering in the presence 
of a Coulomb potential. Although they did not establish the corresponding power
counting, they did show that the resulting EFT was equivalent to a distorted-wave
or ``modified" effective-range expansion \cite{bj,bethe,vhk}. Similar ideas have
also been applied to scattering with a one-pion exchange potential as the 
long-range interaction \cite{sf,krthesis}.

In the present work, we use the RG to study two-body scattering in a system with 
a combination of known long-range and unknown short-range forces. To do this, we 
apply a cut-off to the basis of distorted waves (DW's) \cite{newton} for the known 
long-range interaction. Applying the RG as outlined above, we identify the possible
fixed points of the short-range interaction and establish the corresponding power 
counting rules for perturbations around these. If a nontrivial fixed point exists,
the terms in the resulting EFT can be directly related to those in a DW 
effective-range expansion.

In order to implement this treatment, it is essential that one has identified all 
the important low-energy scales for the system. In the case of the Coulomb 
potential, the Bohr radius provides an additional low-energy scale (apart
from the momenta of the particles). So long as this potential is relatively weak, 
there will be a clear separation of scales between it and the short-range strong 
interaction.  By expressing the Bohr radius in units of the cut-off, we ensure that
the Coulomb potential becomes scale independent when we rescale the theory. This 
means that it can be regarded as part of any fixed point of the RG and so its 
effects should be treated to all orders.

More generally our approach can be applied whenever a long-range potential can be
rescaled to make it scale independent. This potential can then be iterated to all 
orders to generate the basis of DW's which forms the starting point for our RG. 
The fixed points of this RG are then the short-range interactions which, in
combination with the long-range one, lead to scale-free behaviour in the limit 
where all low-energy scales are taken to to zero.

Pion-exchange interactions are of particular interest. These introduce another 
scale: the pion mass. Clearly this should be treated as a low-energy scale if one 
wishes to study nucleon-nucleon scattering with momenta comparable to the pion 
mass. Two schemes for incorporating these forces in EFT's for nucleon-nucleon
scattering can be found in the literature. One, proposed by KSW \cite{ksw2,ksw3}, 
treats these forces perturbatively. The other, suggested by Weinberg \cite{wein2} 
and further developed by van Kolck \cite{orvk,vk1,vk4}, iterates these forces to 
all orders.

The KSW scheme is just an extension of the KSW power counting around the 
effective-range fixed point, in which the pion mass provides the only additional
low-energy scale. All of the pion-exchange terms in the rescaled potential
vanish as $\Lambda\rightarrow 0$ and so can be treated perturbatively. This scheme
is consistent with ChPT as implemented in the zero- and one-nucleon sectors.
However, it does seem to be at best slowly convergent in the $s$-wave 
channels \cite{sf,ch,krthesis,fms}.

The alternative scheme of Weinberg and van Kolck (WvK) uses Weinberg power 
counting to organise the potential, the leading terms of which are then resummed 
to all orders.\footnote{Note that we are careful to distinguish here between 
Weinberg power counting for the terms in the potential and the scheme based on
it for treating pion exchange, which was proposed by Weinberg and developed
by van Kolck.} This method has been applied with some success \cite{orvk,vk4,egm}, 
although questions have been raised about its consistency with the usual chiral 
expansion of ChPT \cite{ksw1,bbsvk}. In our present treatment, we shall see that 
this WvK scheme corresponds to identifying an additional low-energy scale in the 
strength of the one-pion-exchange potential.

The outline of this paper is as follows. In Sec.~\ref{sec:rgshort} we summarise 
the basic results of Ref.~\cite{bmr}. This allows us to set up the basic framework 
of the RG for two-body scattering, and to establish the associated notation we 
use. In particular we discuss the two important fixed points: the trivial one and 
one with a bound state at zero energy. We show how a perturbative expansion around 
each fixed point can be used to determine the corresponding power counting. 

In Sec.~\ref{sec:rglong} we extend these ideas to describe scattering in 
presence of a known long-range potential. Our treatment is based on the 
application of a cut-off to the basis of DW's for that potential. We derive
an RG equations which has a similar form to that obtained in Ref.~\cite{bmr},
but which contains an extra term for each low-energy scale associated with 
the long-range potential. We examine the conditions which that potential must 
satisfy in order to lead to fixed-point behaviour. In order to handle potentials 
which are more singular than $r^{-1}$ as $r\rightarrow 0$, we introduce a second 
regulator. Instead of using a zero-range potential to represent the short-range 
interaction, we use one of finite range, specifically a $\delta$-shell. The
radius of this $\delta$-shell can be thought of a factorisation scale separating 
low-momentum physics, which can be treated in terms of perturbations around a 
fixed point, from nonperturbative short-distance physics. Another difference from 
the case of a pure short-range interaction is that there can be marginal 
perturbations around a fixed point, leading to a potential which evolves 
logarithmically with the cut-off.

This method is applied in Sec.~\ref{sec:ex} to several examples of long-range 
forces. Sec.~\ref{sec:coulomb} revisits the example of scattering by a short-range
interaction in the presence of a Coulomb potential. In this case a nontrivial 
fixed point, if one existed, would have a marginal perturbation leading to a 
logarithmic evolution with cut-off. The same methods can still be used to 
construct a potential where the logarithmic terms have been resummed. The
power counting for perturbations away from this potential is an extended version
of KSW counting and corresponds to the Coulomb DW effective-range expansion.

In Sec.~\ref{sec:yukawa}, we examine $^1S_0$ NN scattering in the presence of
the one-pion exchange. This is a Yukawa potential with the same 
singularity at $r=0$ as the Coulomb potential, and so the RG behaviour is
very similar. We show how one can obtain either the KSW or WvK schemes for
the corresponding EFT, depending on the choice of low-energy scales.

Another important type of long-range interaction is the inverse-square potential. 
This is relevant to scattering in three-body systems, as shown by 
Efimov \cite{efim}. The corresponding RG could therefore help to clarify the 
unusual behaviour found in EFT treatments of such systems \cite{bhvk}. This 
approach can also be applied scattering by a pure short-range potential in
higher partial waves, where the centrifugal barrier has an inverse-square form.
Scattering in the presence of a repulsive inverse-square potential is examined
in Sec.~\ref{sec:repisq} and a nontrivial fixed point is found. Perturbations
around from this point scale with noninteger powers of the cut-off and
correspond to terms in a DW effective-range expansion.  The treatment of the 
attractive inverse-square potential is more involved and will be treated in a 
future work.

\section{RG for short-range forces}\label{sec:rgshort}

Before considering the effects of long-range forces, we first set the scene by
reviewing the main features of the RG treatment of two-body scattering by 
short-range forces \cite{bmr}. For simplicity we consider here only $s$-wave 
scattering, although the extension to higher partial waves is straightforward. 

The fully off-shell amplitude $T(k',k,p)$ for scattering of two heavy particles of
mass $M$ by a potential $V(k',k,p)$ satisfies the Lippmann-Schwinger (LS) equation 
\cite{newton}
\begin{equation} \label{eq:lset}
T(k',k,p)=V(k',k,p)+\frac{M}{2\pi^2}\int q^2{\rm d}q\,\frac{V(k',q,p)T(q,k,p)}
{{p^2}-{q^2}+i\epsilon}.
\end{equation}
Here, as throughout this paper, we use $k$ and $k'$ to denote relative momenta 
and the energy-dependence is expressed in terms of $p\equiv\sqrt{ME}$, the on-shell 
momentum corresponding to the centre-of-mass energy $E$. The on-shell amplitude 
$T(p,p,p)$ is related to the phase shift $\delta(p)$ by
\begin{equation}
T(p,p,p)=-\frac{4\pi}{M}\frac{1}{p\cot\delta(p)-ip}.
\end{equation}

For systems where the scattering is weak at low energies, we can expand the 
on-shell amplitude as a power series in $p$. However for nucleon-nucleon 
scattering, and other systems with bound states or resonances close to 
threshold, such an expansion is convergent only for a very limited range of 
energies. In such cases it more useful to expand the inverse of the amplitude
in the form of an effective-range expansion (ERE) \cite{ere,bj,bethe,newton}
\begin{equation} \label{eq:ere}
-\frac{4\pi}{M}\frac{1}{T(p,p,p)}+ip=p\cot\delta(p)
=-\frac{1}{a}+\frac{1}{2}r_{e}p^2+\cdots,
\end{equation}
where $a$ is the scattering length and $r_e$ is the effective range.

\subsection{RG equation}

At low energies, where the wavelengths of the particles are large compared to
the range of the forces, these forces may be represented by an effective
Lagrangian or Hamiltonian that consists of  contact interactions only. 
In coordinate space these interactions can be expressed as $\delta$-functions
and derivatives of $\delta$-functions. In general these will include time
derivatives (in a Lagrangian framework) or energy dependence (in a Hamiltonian).
In this case, the effective potential has the momentum-space form 
\begin{equation} \label{eq:pot1}
V(k',k,p)=c_{00}+c_{20}(k^2+k^{\prime 2})+c_{02}\,p^2+\cdots,
\end{equation}
to second order in the momentum expansion. For an EFT to be truly effective, we
need to be able to organise the terms in this potential in a systematic way,
according to some power counting. The RG provides a framework for doing this.

In order to derive an RG equation for the potential, it is convenient to 
start from the reactance matrix, $K$. The off-shell $K$-matrix satisfies a
Lippmann-Schwinger (LS) equation that is very similar to that for the
scattering matrix $T$, Eq.~(\ref{eq:lset}), except that the Green's function 
obeys standing-wave boundary conditions. This means that $K$ is real
below all thresholds for particle production. On-shell the $K$- and 
$T$-matrices are related by
\begin{equation}\label{eq:kt}
\frac{1}{K(p,p,p)}=\frac{1}{T(p,p,p)}+\frac{iM p}{4\pi}.
\end{equation}

The LS equation (\ref{eq:lset}) for scattering by contact interactions 
contains loop integrals that diverge and so need to be regularised. This can be 
done using a cut-off \cite{vk3} or dimensional regularisation \cite{ksw1,ksw2}. It 
is also possible to renormalise the equation without specifing a particular
regularisation by making a subtraction \cite{geg}. As the first step in deriving a 
Wilsonian RG equation, we choose to impose a sharp cut-off at $q=\Lambda$ on the 
relative momentum $q$
in the loop. The LS equation for the $K$-matrix is then
\begin{equation} \label{eq:lsek}
K(k',k,p)=V(k',k,p)+\frac{M}{2\pi^2}{\cal{P}}
\int_{0}^{\Lambda}q^2{\rm d}q\,\frac{V(k',q,p)K(q,k,p)}
{{p^2}-{q^2}},
\end{equation}
where ${\cal P}$ denotes the principal value of the integral. 

The RG equation for the effective potential is obtained by allowing $V$ to
depend on the cut-off $\Lambda$ and demanding that the off-shell $K$-matrix be 
independent of $\Lambda$. This is obviously sufficient to ensure that all 
scattering observables do not depend on $\Lambda$. The stronger condition on 
the off-shell $K$-matrix is needed in order to derive the RG equation.
Schematically, the LS equation can be written as 
\begin{equation}\label{eq:lseks}
K=V(\Lambda)+V(\Lambda)G_0(\Lambda)K,
\end{equation}
where $G_0(\Lambda)$ is the regulated free Green's function in Eq.~(\ref{eq:lsek}). 
Differentiating this with respect to $\Lambda$ and demanding that $K$ be 
independent of $\Lambda$ gives
\begin{equation}
\frac{\partial V}{\partial\Lambda}\Bigl(1+G_0(\Lambda)K\Bigr)
+V(\Lambda)\frac{\partial G_0}{\partial\Lambda}K=0.
\end{equation}
Finally, by operating from the right with $(1+G_0K)^{-1}$ and using the fact 
that $K$ satisfies the LS equation (\ref{eq:lseks}), we get
\begin{equation}\label{eq:urge}
\frac{\partial V}{\partial\Lambda} 
=\frac{M}{2\pi^2}V(k',\Lambda,p,\Lambda)\frac{\Lambda^2}{\Lambda^2-p^2}
V(\Lambda,k,p,\Lambda).
\end{equation}

The solution to this equation should satisfy boundary conditions that follow 
from the fact that the effective potential is to describe low-energy scattering 
by short-ranged interactions. For $s$-wave scattering, it should be an analytic 
function of $k^2$ and $k^{\prime 2}$ for small $k$ and $k'$. Rotational invariance 
would also allow dependence on ${\bf k}\cdot{\bf k}'$, but this is needed only 
for higher partial waves. If the energy lies below all thresholds for 
production of other particles then the potential should also be an analytic 
function of the energy, $E=p^2/M$. Under these restrictions, we see that
$V$ should have an expansion in non-negative, integer powers of $k^2$, 
$k^{\prime 2}$ and $p^2$.

The second step in obtaining a Wilsonian RG equation is to rescale all
dimensioned quantities, expressing them in terms of the cut-off $\Lambda$.
Dimensionless momentum variables are defined by $\hat k=k/\Lambda$ etc.,
along with a rescaled potential,
\begin{equation}\label{eq:scale}
\hat V(\hat k',\hat k,\hat p,\Lambda)=\frac{M\Lambda}{2\pi^2}
V(\Lambda\hat k',\Lambda\hat k,\Lambda\hat p,\Lambda),
\end{equation}
where the factor of $M$ corresponds to dividing out $1/M$ from the whole
Hamiltonian before rescaling. In terms of these rescaled quantities,  
Eq.~(\ref{eq:urge}) becomes
\begin{equation}\label{eq:rge}
\Lambda\frac{\partial\hat V}{\partial\Lambda}
=\hat k'\frac{\partial\hat V}{\partial\hat k'}
+\hat k\frac{\partial\hat V}{\partial\hat k}
+\hat p\frac{\partial\hat V}{\partial\hat p}
+\hat V+\hat V(\hat k',1,\hat p,\Lambda)\frac{1}{1-\hat p^2}
\hat V(1,\hat k,\hat p,\Lambda).
\end{equation}
This is the RG equation for the effective potential.

A systematic expansion of the effective potential can be found if, as the 
cut-off is taken to zero, the potential tends to a fixed point of the RG. The 
fixed points are the solutions of Eq.~(\ref{eq:rge}) that do not depend on 
$\Lambda$. Two such points are of particular interest and these are described 
below.

\subsection{Trivial fixed point}

The simplest example of a fixed-point solution to Eq.~(\ref{eq:rge}) is the 
trivial one:
\begin{equation}
\hat V(\hat k',\hat k,\hat p,\Lambda)=0.
\end{equation} 
The $K$-matrix for this potential is also zero, corresponding to no scattering.
This obviously describes a scale-free system.

For systems where the scattering at low energies is weak, the rescaled 
potential tends towards this fixed point as we lower the cut-off towards zero. 
In such cases we can describe the low-energy behaviour in terms of perturbations 
around the trivial fixed point. We can find perturbations which scale with 
definite powers of $\Lambda$ by linearising the RG equation (\ref{eq:rge}) 
about the fixed point and looking for solutions of the form 
\begin{equation}
\hat V(\hat k',\hat k,\hat p,\Lambda)=C\Lambda^\nu \phi(\hat k',\hat
k,\hat p),
\end{equation}
where the functions $\phi$ satisfy the eigenvalue equation
\begin{equation}\label{eq:linrge.tr}
\hat k'\frac{\partial\phi}{\partial\hat k'}
+\hat k\frac{\partial\phi}{\partial\hat k}
+\hat p\frac{\partial\phi}{\partial\hat p}
+\phi=\nu\phi.
\end{equation}
The solutions to this which are well-behaved as the momenta and energy tend to 
zero are simply
\begin{equation}
\phi(\hat k',\hat k,\hat p)=\hat k^{\prime 2l}\hat k^{2m} \hat p^{2n},
\end{equation}
with RG eigenvalues $\nu=2(l+m+n)+1$, where $l$, $m$ and $n$ are non-negative 
integers. The RG eigenvalues are all positive and so the fixed point is a
stable one: starting from any potential in the vicinity of $\hat V=0$ the
RG flow will take the potential to the fixed point as $\Lambda\rightarrow 0$.

The potential near the trivial fixed point can be written in terms of these
perturbations as
\begin{equation}\label{eq:potexp.tr}
\hat V(\hat k',\hat k,\hat p,\Lambda)=\sum_{l,m,n} C_{lmn}
\Lambda^\nu \hat k^{\prime 2l}\hat k^{2m} \hat p^{2n}.
\end{equation}
This potential is Hermitian if we take real coefficients with
$C_{lmn}=C_{mln}$. Note that terms with forms like $i(k^2-k^{\prime 2})$ need
not be included since they vanish after integration by parts in coordinate 
space. The unscaled form of this potential is
\begin{equation}\label{eq:potexp.utr}
V(k',k,p,\Lambda)=\frac{2\pi^2}{M}\sum_{l,n,m}C_{lmn}
k^{\prime 2l}k^{2m} p^{2n}.
\end{equation}
The coefficients in this expansion can be written in a dimensionless form by 
pulling out a factor of $\Lambda_0^{-\nu}$, where $\Lambda_0$ is some scale 
associated with the underlying physics. In a ``natural" theory it is possible to 
choose $\Lambda_0$ so that the dimensionless coefficients are all of order unity.
This scale $\Lambda_0$ then determines where our expansion of the potential breaks 
down. 

The RG eigenvalues $\nu$ provide a systematic way to classify the terms in
this potential, those with the smallest values of $\nu$ dominating for small 
$\Lambda$. Alternatively, in unscaled units, we see that the terms in the
potential can be classified according to the powers of momenta and energy they
contain. For the trivial fixed point, this power counting is just the one 
proposed by Weinberg \cite{wein2}, where an order $d=\nu-1$ is assigned to each 
term.

The behaviour near this fixed point can be used to describe systems 
where the scattering length is small, and so the scattering at low energies 
can be treated perturbatively. For such systems, this power counting has been
shown to provide a systematic treatment in the context of both dimensional 
regularisation with minimal subtraction \cite{ksw1} and cut-off 
approaches \cite{bcp}. The corresponding $K$-matrix is just given by the first 
Born approximation for the unscaled form of the potential 
Eq.~(\ref{eq:potexp.tr}) \cite{bmr}. This is because higher-order terms from
the LS equation are cancelled by higher-order terms in the potential from the
full, nonlinear RG equation. On-shell we have
\begin{equation}
K(p,p,p)=\frac{2\pi^2}{M}\left[C_{000}+\left(C_{001}+C_{010}+C_{100}\right)p^2
+\cdots\right],
\end{equation}
showing that the $C_{lmn}$ are directly related to scattering observables, namely 
the coefficients in the expansion of the $K$-matrix in powers of the energy.

However, one should note that terms of the same order in the energy, $p^2$, or 
momenta, $k^2$ or $k^{\prime 2}$, are equivalent. It is thus possible to swap between 
energy or momentum dependence in the potential without affecting physical 
observables. This can be done by a unitary transformation on the wave function 
or, in field theoretic language, by a transformation of the field variables (see, 
for example: \cite{fear,fht}). Since this involves the combination $p^2-k^2$, which 
vanishes on-shell, it is sometimes referred to as ``using the equations of motion" 
to eliminate either energy or momentum dependence.

\subsection{Effective-range fixed point}

The next simplest fixed-point potential $\hat V=\hat V_0(\hat p)$ depends on 
energy, but not on momenta. It satisfies the RG equation
\begin{equation}
\hat p\frac{{\rm d}\hat V_0}{{\rm d}\hat p}
+\hat V_0+\frac{\hat V_0^2}{1-\hat p^2}=0.
\end{equation}
A convenient way to solve this equation, as well as other momentum-independent RG 
equations, is to divide it by $\hat V_0^2$ to obtain a linear equation for
$1/\hat V_0$:
\begin{equation}
\hat p\frac{{\rm d}}{{\rm d}\hat p}\left(\frac{1}{\hat V_0}\right)
-\frac{1}{\hat V_0}-\frac{1}{1-\hat p^2}=0.
\end{equation}
This equation is satisfied by the basic loop integral
\begin{equation}
\hat J(\hat p)={\cal P}\int_0^1 \frac{\hat q^2{\rm d}\hat q}{\hat p^2-\hat q^2}
=-1+\frac{\hat p}{2}\ln{\frac{1+\hat p}{1-\hat p}}.
\end{equation}
This is analytic in $\hat p^2$ as $\hat p^2\rightarrow 0$ and so it gives us
the fixed-point potential
\begin{equation}
\hat V_0(\hat p)=-\left[1-\frac{\hat p}{2}\ln{\frac{1+\hat p}{1-\hat p}}\right]^{-1}.
\end{equation}

When this potential used in the LS equation (\ref{eq:lsek}), the corresponding 
$K$-matrix is found to give an infinite scattering length, 
\begin{equation}
\frac{1}{K(p)}=0.
\end{equation}
Alternatively one can think of it as generating a bound state at zero energy.
Since the energy of the bound state is exactly zero, the system has no scale
associated with it, which is why it is described by a fixed point of the RG.

This fixed point is the one of most interest for low-energy nuclear physics.
Systems such as two nucleons, which have $s$-wave bound states or resonances near 
threshold, can be described by potentials close to this fixed point. Such a 
potential can be expanded in terms of small perturbations away from $\hat V_0$
that scale with definite powers of $\Lambda$: 
\begin{equation}
\hat V(\hat k',\hat k,\hat p,\Lambda)=\hat V_0(\hat p)
+C\Lambda^\nu \phi(\hat k',\hat k,\hat p).
\end{equation}
These perturbations satisfy the linearised RG equation
\begin{equation}\label{eq:linrge.ere}
\hat k'\frac{\partial\phi}{\partial\hat k'}
+\hat k\frac{\partial\phi}{\partial\hat k}
+\hat p\frac{\partial\phi}{\partial\hat p}+\phi
+\frac{\hat V_0(\hat p)}{1-\hat p^2}\left[
\phi(\hat k',1,\hat p)+\phi(1,\hat k,\hat p)\right]
=\nu\phi.
\end{equation}

Of particular interest are the eigenfunctions of Eq.~(\ref{eq:linrge.ere}) that 
depend only on energy. These have the form
\begin{equation}\label{enper.ere}
\phi(\hat p)=\hat p^{\nu+1} \hat V_0(\hat p)^2.
\end{equation}
If we demand that they be well-behaved functions of $p^2$ as $p^2\rightarrow 
0$, then we find that the allowed RG eigenvalues are $\nu=-1,1,3,\dots$. The 
presence of one negative eigenvalue shows that the fixed point is unstable.

Only potentials which lie on the ``critical surface" where the coefficient of
the unstable eigenvector is zero flow to the nontrivial fixed point as 
$\Lambda\rightarrow 0$. All others go either to the trivial fixed point or to
infinity. Potentials which lie close to the cricial surface will flow
initially towards the nontrivial fixed point and so their behaviour can be
analysed using the expansion around this point. Eventually however, when $\Lambda$ 
becomes small enough, their flows are pushed away by the unstable perturbation. This
behaviour is illustrated below.

\begin{figure}
\includegraphics*[80,510][420,720]{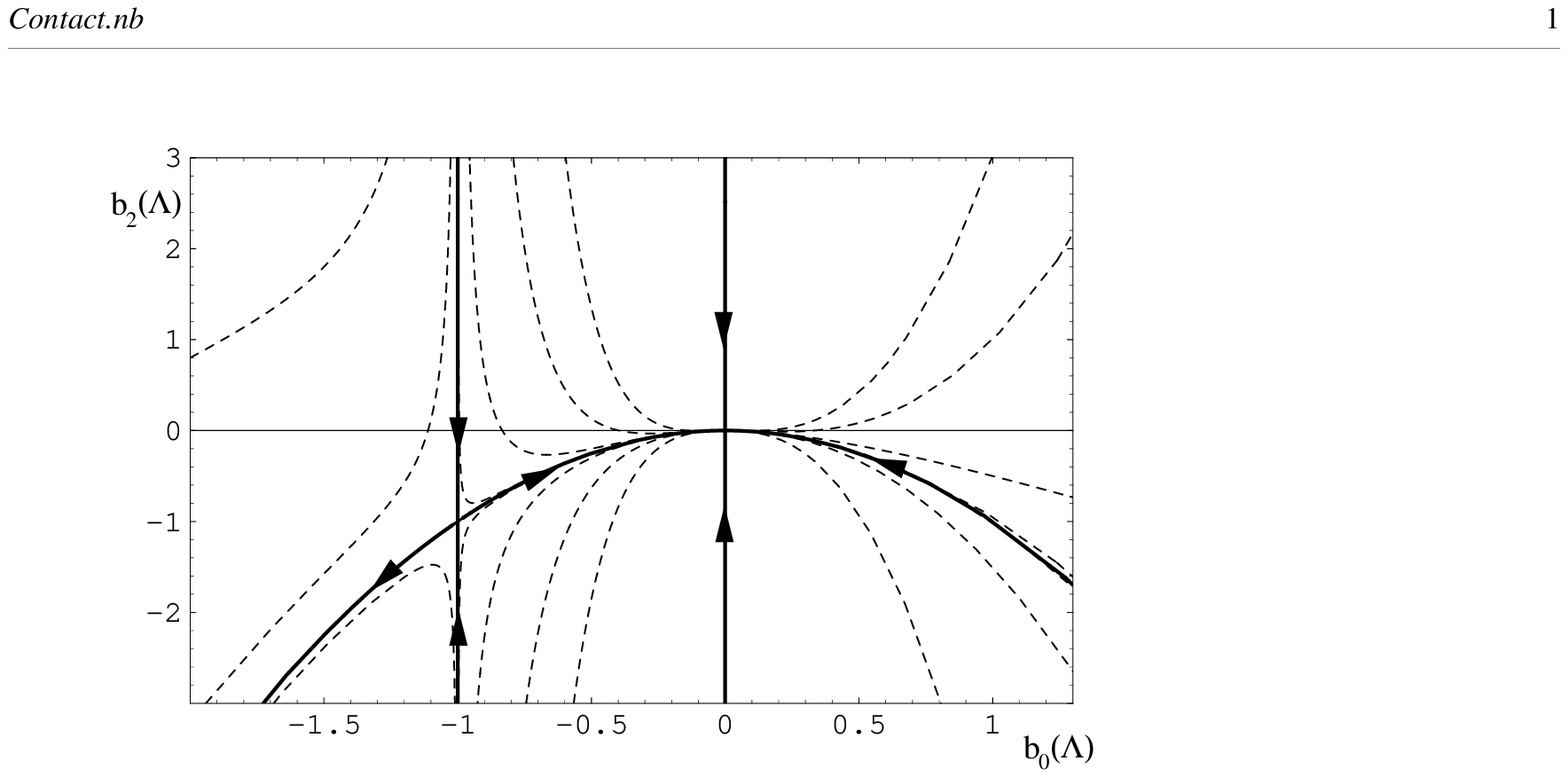}
\caption{The flow as $\Lambda\rightarrow 0$ of the first two coefficients in the 
expansion Eq.~(\ref{eq:enexp}) of the rescaled potential.}
\label{fig:SRFlow}
\end{figure}

A useful method of examining RG flows is shown in Fig.~\ref{fig:SRFlow}. This 
illustrates the flow as $\Lambda\rightarrow 0$ in a two-dimensional slice of the 
infinite-dimensional space of effective potentials. In particular it shows the 
behaviour of the first two coefficients in the expansion of the rescaled potential
in powers of the energy,
\begin{equation}\label{eq:enexp}
\hat V=b_0(\Lambda)+b_2(\Lambda)\hat p^2+\cdots.
\end{equation}
There are two fixed points: the trivial one located at the origin and the nontrivial
one at $(-1,-1)$. The flow lines in bold lie along RG eigenvectors close to the 
fixed points; the dashed lines show more general flow lines. All potentials near 
the trivial fixed point flow into it, and so that point is stable. In contrast only
potentials lying on the critical surface defined by the stable perturbations (the
vertical line $b_0=-1$ in the figure) flow to the nontrivial fixed point.
We see that flows near the critical surface are governed by the power counting 
around this fixed point until $\Lambda$ reaches some critical scale when the 
unstable perturbation kicks in and takes the flow either into the domain of the 
trivial fixed point or to infinity.  We shall see below that this critical scale is 
in fact the inverse of the scattering length.

The power counting for perturbations about this fixed point is different from
that for the trivial fixed point. The energy-dependent perturbations can be 
assigned orders $d=\nu-1=-2,0,2,\dots$ in this scheme, which is the one also 
obtained by Bedaque and van Kolck \cite{bvk1,vk3}, Kaplan, Savage and 
Wise \cite{ksw2}, and Gegelia \cite{geg}. An (unscaled) potential close to the fixed 
point can be expanded in terms of these perturbations as
\begin{equation}\label{eq:pot.ere}
V(p,\Lambda)=V_0(p,\Lambda)+\frac{M}{2\pi^2}\sum_{n=0}^\infty C_{2n-1}p^{2n}
V_0(p,\Lambda)^2.
\nonumber
\end{equation}

The connection with scattering observables can be found by solving the LS equation 
for this potential. The resulting on-shell $K$-matrix is given by
\begin{equation}
\frac{1}{K(p)}=-\frac{M}{2\pi^2}\sum_{n=0}^{\infty}
C_{2n-1}p^{2n}.
\end{equation}
By comparing this with the ERE, Eq.~(\ref{eq:ere}), we see 
that there is a one-to-one correspondence between the perturbations around the 
nontrivial fixed point and the terms in the effective range expansion.
In particular, the first two coefficients are
\begin{equation}\label{eq:erecfs}
C_{-1}=-\frac{\pi}{2a},\qquad C_1=\frac{\pi r_e}{4}.
\end{equation}
In $s$-wave nucleon-nucleon scattering, the scattering lengths, $a$, are large. 
This implies that the coefficient $C_{-1}$ can be treated as a small perturbation. 
In this case, as explained above, the effective potential remains close to the 
critical surface flowing into the nontrivial fixed point, provided the cut-off
$\Lambda$ is kept larger than $1/a$. The power counting about this fixed point
is therefore the appropriate one for organising the terms in the nucleon-nucleon
potential for momenta between $1/a$ and the pion mass.

Finally we note that there are also momentum-dependent perturbations. The
RG analysis of Ref.~\cite{bmr} showed that these have even RG eigenvalues.
They are thus of odd orders in the power counting about the nontrivial 
fixed point and they do not contribute to on-shell observables. For example, the 
perturbation with eigenvalue $\nu=2$ has the form (in physical units)
\begin{equation}
C_2\left\{\left[k^2-p^2+\frac{1}{3}\frac{M\Lambda^3}{2\pi^2}
V_0\right]V_0(p,\Lambda)+\;(k\rightarrow k')\right\}.
\end{equation} 
This consists of pieces which vanish on-shell, together with a term that cancels
a cubic divergence generated by $k^2$ terms. Note that energy and momentum
dependence are not equivalent in the vicinity of the nontrivial fixed point.
The fact that they appear at different orders in the expansion means that it is
not possible to generate a nonzero effective range from a purely momentum-dependent
potential without an unnaturally large coefficient. A similar point about the
need to include energy dependence has also been noted by Beane and 
Savage \cite{bs}.

\section{RG with long-range forces}\label{sec:rglong}

Now consider a system of two particles interacting through a potential
\begin{equation}
V=V_L+V_S,
\end{equation}
where $V_L$ is a known long-range potential and $V_S$ contains the effects
of short-range physics, which we wish to parametrise systematically. 

To obtain an RG equation for the short-range potential alone, it is convenient
to work in terms of the distorted waves (DW's) of the long-range potential. The 
$T$-matrix describing scattering by $V_L$ alone is 
\begin{equation}
T_L=V_L+V_L G_0^+ T_L,
\end{equation}
and the corresponding Green's function is
\begin{equation}
\hat G_L^+=G_0^++G_0^+ T_L G_0^+.
\end{equation}
By using the ``two-potential trick" to resum the effects of $V_L$ to all
orders \cite{newton}, the full $T$-matrix can be written in the form
\begin{equation}\label{eq:dwt}
T=T_L+(1+T_L G_0^+)\tilde T_S (1+ G_0^+ T_L),
\end{equation}
where $\tilde T_S$ satisfies the LS equation
\begin{equation}\label{eq:dwlst}
\tilde T_S=V_S+V_S G_L^+ \tilde T_S.
\end{equation}
The expression (\ref{eq:dwt}) for $T$ is the starting point for the distorted-wave 
Born approximation \cite{newton} since $\Omega=1+G_0^+ T_L$ is the M\o ller wave 
operator which converts a plane wave into a DW of $V_L$. The operator $\tilde T_S$ 
therefore describes the scattering between the DW's as a result of the 
short-range potential. 

The effects of the short-range potential can be expressed in terms of 
\begin{equation}
\tilde\delta_S=\delta-\delta_L,
\end{equation}
the difference between the full phase shift, $\delta$, and that due to $V_L$ alone,
$\delta_L$. By taking on-shell matrix elements of Eq.~(\ref{eq:dwt}), the matrix
elements of $\tilde T_S$ between DW's can be written in the form
\begin{equation}\label{eq:dwtps}
\langle \psi_L^-(p)|\tilde T_S(p)|\psi_L^+(p)\rangle
=-\frac{4\pi}{M}e^{2i\delta_L(p)}\frac{e^{2i\tilde\delta_S(p)}-1}{2ip},
\end{equation}
where $\psi_L^+(p,r)$ is the outgoing DW of $V_L$ with energy $E=p^2/M$ and 
$\psi_L^-$ is the corresponding incoming wave (and for simplicity we are 
considering only $s$-wave scattering). Bethe \cite{bethe} used this as the 
starting point for constructing a ``modified" or distorted-wave effective-range 
expansion (DWERE). His result can be written in the form
\begin{eqnarray}\label{eq:dwere}
&&\displaystyle{e^{2i\delta_L(p)}|\psi_L^+(p,0)|^2
\left(\frac{1}{\langle\psi_L^-|\tilde T_S(p)
|\psi_L^+\rangle}-ip\right)+{\cal M}_L(p)}\nonumber\\
&&\displaystyle{\qquad\qquad=|\psi_L^+(p,0)|^2\,p\,(\cot\tilde\delta_S(p)-i)
+{\cal M}_L(p)=-\frac{1}{\tilde a}+\frac{1}{2}\tilde r_ep^2+\cdots.}
\end{eqnarray}
where the function ${\cal M}_L(p)$ is the logarithmic derivative
at the origin of the Jost solution to the Schr\"odinger equation with
the potential $V_L$ \cite{vhk}. This expansion has long been used to extract 
low-energy properties of the strong interaction between two 
protons \cite{bethe,jb,vhk,kr} and more recently to remove the effects of 
one-pion exchange from NN scattering \cite{sf,krthesis}.

The DWERE is useful because all rapid, and possibly
nonanalytic, dependence on the energy is removed in ${\cal M}_L(p)$ and
$e^{2i\delta_L(p)}|\psi_L^+(p,0)|^2$. This leaves an amplitude which can be
expanded as a power series in the energy, with coefficients whose scale is set
by the underlying short-distance physics. Together with the appearance of wave 
functions at the origin, this strongly suggests that it
should be possible to re-express this expansion in the form of an EFT. Indeed 
Kong and Ravndal \cite{kr} have done just this for the case of the Coulomb 
potential, but without explicitly establishing the form of the corresponding 
power counting.

\subsection{RG equation}

The power counting which makes possible a systematic expansion of the 
short-range potential can be obtained from an RG analysis which is similar to 
that in Sec.~\ref{sec:rgshort} but which incorporates the distortion effects of 
the long-range potential. Such a DW version of the RG is most conveniently 
obtained from the DW $K$-matrix which satisfies an LS equation similar to
Eq.~(\ref{eq:dwlst}) but with standing-wave boundary conditions,
\begin{equation}\label{eq:dwlsk}
\tilde K_S=V_S+V_SG_L\tilde K_S.
\end{equation}
To regulate this equation, we apply a cut-off to $G_L$ in the DW basis,
\begin{equation}
G_L=\frac{M}{2\pi^2}{\cal P}\int_0^\Lambda {\rm d}q\,q^2\frac{|\psi_L(q)\rangle
\langle\psi_L(q)|}{p^2-q^2}\;(+\hbox{bound states}).
\end{equation}
If we demand that $\tilde K_S$ should be independent of the cut-off $\Lambda$,
then we can use the same arguments as above to find that the short-range 
potential must satisfy the equation
\begin{equation}\label{eq:udwrge}
\frac{\partial V_S}{\partial\Lambda}=-V_S\frac{\partial G_L}{\partial\Lambda}V_S.
\end{equation}
If we did not apply the cut-off to the DW basis, $V_S$ would have to
change with $\Lambda$ not only to incorporate the physics that has been
integrated out but also to correct for the modification of $V_L$ by the
cut-off. The resulting equation for $V_S$ would then have taken a much more 
complicated form.

To simplify the analysis, we assume here that $V_S$ depends only on energy, not
on momenta. As shown in Ref.~\cite{bmr}, the momentum-dependent solutions to the 
RG equation have more complicated forms but are not needed to describe on-shell 
scattering. In Sec.~\ref{sec:rgshort} the short-range force was taken to be a 
contact interaction, $V_S(p,\Lambda;r)=V_S(p,\Lambda)\delta(\bf{r})$. This is the 
simplest form that satisfies the appropriate symmetries for $s$-wave scattering. 
However such a choice cannot be used in combination with most of the long-range
potentials of interest. These potentials are sufficiently singular that their 
DW's $\psi_L(p,r)$ either vanish or diverge as $r\rightarrow 0$. As a result the 
right-hand side of Eq.~(\ref{eq:udwrge}) is either zero or ill-defined for a contact 
interaction.

Instead of a contact interaction we chose a spherically symmetric potential 
with a short but nonzero range. By choosing the range $R$ of this potential to be
much smaller than $1/\Lambda$, we ensure that any additional energy or 
momentum dependence associated with it is no larger than that of the physics 
which has been integrated out, and hence the power-counting is not altered by it.
This scale $R$ separates the region of low-momentum physics, which we wish 
describe with an EFT, from a region of high momentum physics, which is 
nonperturbative as a result of the singular behaviour of the long-range
potential. The precise value of this scale is arbitrary and so observables
should not depend on it.

A simple and convenient choice for the form of this regulator is the 
``$\delta$-shell" potential,
\begin{equation}
V_S(p,\kappa,\Lambda;r)=V_S(p,\kappa,\Lambda)\frac{\delta(r-R)}{4\pi R^2},
\end{equation}
where $\kappa$ denotes a generic low-energy scale associated with $V_L$. 
With this choice, the equation (\ref{eq:udwrge}) for $V_S$ becomes
\begin{equation}\label{eq:udwrge2}
\frac{\partial V_S}{\partial\Lambda}=-\frac{M}{2\pi^2}
|\psi_L(\Lambda,R)|^2\frac{\Lambda^2}{p^2-\Lambda^2}
V_S^2(p,\kappa,\Lambda),
\end{equation}
where $\psi_L(k,R)$ is the DW with asymptotic momentum $k$,
evaluated at $r=R$. This DW satisfies the Schr\"odinger equation,
\begin{equation}\label{eq:Schrodinger}
\left(\frac{{\rm d}^2}{{\rm d}r^2}+\frac{2}{r}\frac{\rm d}{{\rm d}r}\right) 
\psi_L(p,r)-MV_L(r,\kappa)\psi_L(p,r)+p^2\psi_L(p,r)=0,
\end{equation}
and is normalised so that
\begin{equation}
\int_0^\infty {\rm d}r\,p^2 r^2\psi_L(p,r)\psi_L(p',r)=\frac{\pi}{2}\delta(p-p').
\end{equation}

In order to include nonperturbatively all of the physics associated with the 
long-range potential we must rescale $\kappa$ and keep $\hat\kappa=\kappa/\Lambda$ 
fixed as we take $\Lambda\rightarrow 0$. If, when we rescale $V_L$ as in 
Eq.~(\ref{eq:scale}), the potential is independent of $\Lambda$, we can  
regard it as part of a fixed point. This ensures that the long-range potential 
continues to influence the fixed points of the effective potential and the 
long-range physics is not ``integrated out". If the rescaled $V_L$ vanishes
as $\Lambda\rightarrow 0$, we can treat it as a small perturbation about 
one of the fixed points of the pure short-range potential. Finally, if $V_L$
grows as $\Lambda\rightarrow 0$, then it is not possible to find any fixed-point
behaviour. In such cases we must either give up the RG analysis or try to identify 
additional low-energy scales in the problem. 

The RG equation is obtained in the same way as before, by rescaling $p$, $\kappa$ 
and $V_S$ in Eq.~(\ref{eq:udwrge2}). The appearance of the factor 
$|\psi_L(\Lambda,R)|^2$ in this equation means that the effective potential must
be rescaled in a way that depends on the long-range potential. We assume that the 
wave function in the limit $R\ll 1/p$ takes the separable form
\begin{equation}\label{eq:wfform}
|\psi_L(p,R)|^2=|{\cal N}(\kappa/p)|^2 f(p)F(R).
\end{equation} 
This is general enough to cover all the examples studied here and also the attractive
inverse-square potential. Here ${\cal N}(\kappa/p)$ is a normalisation factor 
through which information about the long-distance physics is communicated to the 
short-distance physics. 

To remove the dependence on $\Lambda$ and $R$ contained in $|\psi_L(\Lambda,R)|^2$ 
from the right-hand side of Eq.~(\ref{eq:udwrge2}), we define the rescaled 
short-range potential 
\begin{equation}\label{eq:dwscale}
\hat V_S(\hat p,\hat\kappa,\Lambda)=\frac{M \Lambda}{2 \pi^2}F(R)f(\Lambda)
V_S(\Lambda\hat p,\Lambda\hat\kappa,\Lambda).
\end{equation}
This satisfies the DWRG equation
\begin{equation}\label{eq:dwrge}
\Lambda\frac{\partial\hat V_S}{\partial\Lambda}
=\hat p\frac{\partial\hat V_S}{\partial\hat p}
+\hat\kappa\frac{\partial\hat V_S}{\partial\hat\kappa}
+\left(1+\Lambda\frac{f'(\Lambda)}{f(\Lambda)}\right)\hat V_S
+\frac{|{\cal N}(\hat\kappa)|^2}{1-\hat p^2}\hat V^2_S.
\end{equation}
Since we have chosen to study potentials which are independent of momentum, as
discussed above, the RG equation can be divided by $\hat V_S^2$ to obtain a more 
practical linear equation for $1/\hat V_S$. This equation has the form
\begin{equation}\label{eq:dwrge2}
\Lambda\frac{\partial}{\partial\Lambda}\left(\frac{1}{\hat V_S}\right)
=\hat p\frac{\partial}{\partial\hat p}\left(\frac{1}{\hat V_S}\right)
+\hat\kappa\frac{\partial}{\partial\hat\kappa}\left(\frac{1}{\hat V_S}\right)
-\left(1+\Lambda\frac{f'(\Lambda)}{f(\Lambda)}\right)\frac{1}{\hat V_S}
-\frac{|{\cal N}(\hat\kappa)|^2}{1-\hat p^2}.
\end{equation}

From the renormalization group equation (\ref{eq:dwrge}) we can see that there is
always a trivial fixed-point solution, $\hat V_S=0$. However nontrivial fixed points
can only exist if the right-hand side is independent of $\Lambda$. This occurs if 
the wave function has a power-law form close to the origin, 
\begin{equation}\label{eq:wfpl}
|\psi_L(p,R)|^2=|{\cal N}(\kappa/p)|^2 (p R)^{\sigma-1},
\end{equation} 
where $\sigma$ is a real number. This is the case for all the examples considered 
here, but not for the attractive inverse-square potential. Taking 
$f(\Lambda)=\Lambda^{\sigma-1}$, the DWRG equation can be written
\begin{equation}\label{eq:dwrge3}
\Lambda\frac{\partial}{\partial\Lambda}\left(\frac{1}{\hat V_S}\right)
=\hat p\frac{\partial}{\partial\hat p}\left(\frac{1}{\hat V_S}\right)
+\hat\kappa\frac{\partial}{\partial\hat\kappa}\left(\frac{1}{\hat V_S}\right)
-\sigma\frac{1}{\hat V_S}-\frac{|{\cal N}(\hat\kappa)|^2}{1-\hat p^2}.
\end{equation}

As in the pure short-range case, the effective potential must satisfy the boundary 
condition of being analytic in $p^2$ at low energies. This arises from the 
assumption that all the nonanalytic energy dependence generated by long-range 
physics has been factored out into the DW's. The corresponding condition for the 
dependence on $\kappa$ depends on the nature of that scale. In most cases we need 
to demand only that the effective potential is analytic in $\kappa$. An example is 
the inverse Bohr radius which forms a low-energy scale for the Coulomb potential.
Since this scale is proportional to the fine-structure constant $\alpha$, the
short-distance physics which has been incorporated in the effective potential should 
be analytic in it. In contrast the pion mass is proportional to the square root of 
the strength of the chiral symmetry breaking (the current quark mass in the 
underlying theory, QCD) and hence, as in ChPT, the effective potential should be 
analytic in $m_\pi^2$.

\subsection{Trivial fixed point}\label{ssec:tfp}

As already noted, a trivial fixed point, $\hat V_0=0$, always exists for 
Eq.~(\ref{eq:dwrge3}). The interpretation of this fixed point is the similar to that 
in the pure short range case. It describes a system with no scattering of the
distorted waves of the long-range potential. Perturbations about this point can
be used to describe systems where the short-range interactions provide small
corrections to the scattering by $V_L$. They correspond to an expansion of the
DW $K$-matrix $\tilde K_S$ of Eq.~(\ref{eq:dwlsk}) in powers of energy and $\kappa$.
They have the forms
\begin{equation}
\hat V_S=C_{mn}\Lambda^{m+2n+\sigma}\hat\kappa^m \hat p^{2n}.
\end{equation}
The boundary conditions demand that $m$ and $n$ are non-negative integers. In 
addition, if $\kappa$ is a scale like the pion mass, $m$ must be even.

The stability of the fixed point depends upon the value of $\sigma$, which describes
the power-law behaviour of the DW's near the origin . For positive $\sigma$ the 
fixed point is stable, while for negative $\sigma$ there is one or more unstable 
perturbation, with $m+2n+\sigma<0$. If $\sigma=0$ then the perturbation with $m=n=0$ 
is marginal. A marginal eigenfunction may also exist when $\sigma$ is a negative 
integer, if $m+2n=-\sigma$ is allowed. These marginal eigenfunctions have no 
power-law evolution with $\Lambda$. Instead we expect to find a logarithmic 
dependence which will allow us to classify such perturbations as marginally stable 
or unstable.

\subsection{Nontrivial fixed point}\label{ssec:dwfp}

A nontrivial fixed point of Eq.~(\ref{eq:dwrge3}), if one exists, satisfies the 
equation,
\begin{equation}\label{eq:fpdwrge}
\hat p\frac{\partial}{\partial\hat p}\left(\frac{1}{\hat V_0}\right)
+\hat\kappa\frac{\partial}{\partial\hat\kappa}\left(\frac{1}{\hat V_0}\right)
=\sigma\frac{1}{\hat V_0}+\frac{|{\cal N}(\hat\kappa)|^2}{1-\hat p^2}.
\end{equation}
It is useful to note that this equation for $1/V_0$ is satisfied by the loop integral
\begin{equation}\label{eq:bloop}
\hat J_0(\hat p,\hat\kappa)={\cal P}\int_0^1\hat q^{\sigma+1}{\rm d}\hat q
\frac{|{\cal N}(\hat\kappa/\hat q)|^2}{\hat p^2-\hat q^2}.
\end{equation}
Unlike the pure short-range case, we cannot directly identify $1/\hat V_0=\hat J_0$ 
as a fixed-point solution of the RG equation since it is not in general analytic 
as $\hat p,\hat\kappa\rightarrow 0$. Nonetheless we can use $\hat J_0$ as a starting 
point for finding such solutions. The method will be described in more detail in the
examples below, but the basic idea is to subtract from $\hat J_0$ a solution to the 
homogeneous version of Eq.~(\ref{eq:fpdwrge}) to cancel all its nonanalytic
beaviour. We can then write
\begin{equation}\label{eq:ntfpm}
\frac{1}{\hat V_0}=\hat J_0(\hat p,\hat\kappa)-\hat{\cal M}(\hat p,\hat\kappa),
\end{equation}
where $\hat{\cal M}(\hat p,\hat\kappa)$ is a homogenous function of order
$\sigma$ in $\hat p$ and $\hat\kappa$.

Perturbations around a fixed-point solution to Eq.~(\ref{eq:fpdwrge}) have the form
\begin{equation}\label{eq:ntpert}
\frac{1}{\hat V_S}=\frac{1}{\hat V_0}-C_{mn}\Lambda^{m+2n-\sigma}\hat\kappa^m
\hat p^{2n},
\end{equation}
where $n$ and $m$ satisfy the conditions described above in the case of the trivial 
fixed point. If $\sigma$ is positive then the fixed point is unstable, the 
eigenfunctions with $m+2n-\sigma<0$ corresponding to the unstable directions.
If $\sigma$ is negative then this fixed point is stable. Marginal eigenvectors may 
exist for integer $\sigma$ if $m+2n=\sigma$ is allowed. In particular, for 
$\sigma=0$ the fixed point is marginal with the $m=n=0$ perturbation having zero 
eigenvalue. 

\subsection{Regaining the DW expansion}\label{sec:dwere}

Having obtained a solution to the full RG equation near the nontrivial fixed point,
we can use it in the DW Lippmann-Schwinger equation for $\hat K_S$ (\ref{eq:dwlsk})
in order to connect the potential to scattering observables. As with a pure
short-range interaction, we find a direct connection to an effective range expansion,
in this case the DW version of Eq.~(\ref{eq:dwere}). From Eqs.~(\ref{eq:kt}) and 
(\ref{eq:dwlst}) we see that the on-shell matrix element,
$\langle\psi_L|\hat K_S|\psi_L\rangle$, is related to the additional phase shift,
$\tilde\delta_S$, by
\begin{equation}
\langle\psi_L(p)|\hat K_S|\psi_L(p)\rangle=-\frac{4\pi}{M}\frac{1}
{p\cot \bar\delta_S}.
\end{equation}

The LS equation can be solved by expanding the Green's functions in terms of a 
complete set of DW's and iterating to get a geometric series. This series can then
be summed, giving
\begin{equation}
\langle\psi_L(p)|\hat K_S|\psi_L(p)\rangle
=\frac{V_S(p,\kappa,\Lambda) |\psi_L(p,R)|^2}
{1-\displaystyle{\frac{V_S(p,\kappa,\Lambda) M}{2\pi^2} {\cal P}
\int_0^\Lambda q^2 dq \frac{|\psi_L(q,R)|^2}{p^2-q^2}}}.
\end{equation}  
Note that the integral in the denominator is just 
$\Lambda^\sigma R^{\sigma-1} \hat J_0(\hat p,\hat\kappa)$ where $\hat J_0$ is 
defined in Eq.~(\ref{eq:bloop}).

This equation can be rewritten in terms of the cotangent of the additional phase shift
as
\begin{equation}\label{eq:dwexp1}
|{\cal N}(\kappa/p)|^2 \frac{\pi p^\sigma}{2}\cot\bar\delta_S
=\Lambda^\sigma\left(\hat J_0(\hat p,\hat\kappa)-\frac{1}
{\hat V_S(\hat p,\hat\kappa)}\right).
\end{equation}
This result is independent of $R$, as anticipated. Despite initial appearances it is 
also independent of $\Lambda$. The difference between $1/\hat V_0$ and $\hat J_0$ is
just the homogeneous function $\hat{\cal M}(\hat p,\hat\kappa)$ introduced in 
Eq.~(\ref{eq:ntfpm}). The corresponding piece of the right-hand side is a 
homogeneous function of order $\sigma$ in the physical variables $p$ and $\kappa$. 
Including perturbations of the form given in Eq.~(\ref{eq:ntpert}), we can therefore 
rewrite Eq.~(\ref{eq:dwexp1}) as
\begin{equation}\label{eq:dwexp2}
|{\cal N}(\kappa/p)|^2 \frac{\pi p^\sigma}{2}\cot\bar\delta_S
=-{\cal M}(p,\kappa)+\sum_{n,m}C_{nm}p^n\kappa^m,
\end{equation}
where ${\cal M}(p,\kappa)=\Lambda^\sigma\hat{\cal M}(\hat p,\hat\kappa)$.

This resulting equation (\ref{eq:dwexp2}) has exactly the form of the distorted wave 
expansion, Eq.~(\ref{eq:dwere}). The only difference is that, for sufficiently 
singular potentials, the wave functions must be evaluated close to, but not exactly 
at, the origin. However, provided the wave functions have a power-law form in this 
region, the $R$ dependence cancels leaving Eq.~(\ref{eq:dwexp2}). All nonanalytic 
effects of the long-range force have been factored or subtracted out in the 
functions ${\cal N}$ and ${\cal M}$. The remainder can be written as a power-series 
expansion in $p$ and $\kappa$, which corresponds directly to the expansion of the 
short-range effective potential.

\section{Examples}\label{sec:ex}

\subsection{Coulomb potential}\label{sec:coulomb}

Scattering from a combination of Coulomb and short-range potentials has already 
been studied from an EFT viewpoint by Kong and Ravndal \cite{kr}. Here we examine it 
using the RG and show how a power counting can be established which matches exactly 
with the DWERE.

Since the fine structure constant $\alpha$ is small, a low-energy scale for this
potential is provided by the inverse of the Bohr radius,
\begin{equation}
\kappa=\frac{\alpha M}{2}.
\end{equation}
By rescaling this quantity as described above, we ensure that the long-range potential
remains independent of $\Lambda$ as $\Lambda\rightarrow 0$. In contrast, if we
were to treat $\kappa$ as a high-energy scale, then the Coulomb potential would act 
like a relevant perturbation and would destroy any possible infrared fixed points. 

The correctly normalised distorted $s$-wave for this potential is, in terms of a
hypergeometric function, 
\begin{equation}
\psi_L(p,r)=e^{-\pi\kappa/2p}\Gamma(1-i\kappa/p)\,_1 
F_{1}(1+i\kappa/p;2,-2ipr/\kappa)e^{ipr}.
\end{equation} 
This tends to a finite, nonzero value at the origin and so it has the
form assumed in the previous section with $\sigma=1$. The long-range
phase-shift is $\delta_L={\rm Arg}\Gamma(1+i\kappa/p)$ and the square of the wave 
function at the origin is given by the well-known Sommerfeld factor
\begin{equation}
|{\cal N}(\kappa/p)|^2=\lim_{R\rightarrow 0}|\psi_L(p,R)|^2={\cal C}(\kappa/p)
\equiv\frac{2\pi\kappa/p}{e^{2\pi\kappa/p}-1}.
\end{equation}

A trivial fixed point always exists and the power counting for the expansion around 
it can be found using the general analysis in Sec.~\ref{ssec:tfp}. Of more interest 
are possible nontrivial solutions to the RG and the expansions around them. We take 
the basic 
loop integral of Eq.~(\ref{eq:bloop}) as our a starting point for the construction 
of these solutions. This satisfies Eq.~(\ref{eq:fpdwrge}), the fixed-point version 
of the DW RG equation (\ref{eq:dwrge3}) but it contains nonanalytic terms which 
should not be present. To identify these, we follow Kong and Ravndal \cite{kr} and
break the integral up as
\begin{equation}
\hat J_0(\hat p,\hat\kappa)=-\int_0^1 d\hat q\,{\cal C}(\hat\kappa/\hat q)
+\hat p^2{\cal P}\int_0^\infty d\hat q\frac{{\cal C}(\hat\kappa/\hat q)}{\hat p^2
-\hat q^2}
-\hat p^2\int_1^\infty d\hat q\frac{{\cal C}(\hat\kappa/\hat q)}{\hat p^2-\hat q^2}.
\end{equation}
When written in terms of physical, rather than rescaled, variables, the first term 
contains linear and logarithmic divergences. The second term is finite but depends 
nonanalytically on $\hat p$ and $\hat\kappa$, whereas the final term is analytic.

Using the detailed expression of Kong and Ravndal \cite{kr} for the second term and 
expanding the first in powers of $\hat\kappa$, we can write $\hat J_0$ in the form
\begin{equation}
\hat J_0(\hat p,\hat\kappa)=-1-\pi\hat\kappa\Bigl(\ln\hat\kappa+\gamma\Bigr)
-\pi\hat\kappa{\rm Re}[H(\hat\kappa/\hat p)]+\hbox{terms analytic in 
$\hat p$, $\hat\kappa$},
\end{equation}
where $\gamma$ is Euler's constant and the function $H$ is given by
\begin{equation}
H(x)=\psi(ix)+\frac{1}{2ix}-\ln(ix),
\end{equation}
in terms of the logarithmic derivative of the $\Gamma$-function, denoted by 
$\psi$.

A potential which is analytic as $\hat p^2,\hat\kappa\rightarrow 0$ can be built 
from $\hat J_0$ by cancelling the terms proportional to $\hat\kappa\ln\hat\kappa$ 
and $\hat\kappa{\rm Re}[H(\hat\kappa/\hat p)]$. The second of these satisfies the 
homogeneous version of Eq.~(\ref{eq:fpdwrge}),  
\begin{equation}\label{eq:fphdwrge}
\hat p\frac{\partial}{\partial\hat p}\left(\frac{1}{\hat V_0}\right)
+\hat\kappa\frac{\partial}{\partial\hat\kappa}\left(\frac{1}{\hat V_0}\right)
=\frac{1}{\hat V_0},
\end{equation}
and so it can be subtracted from $\hat J_0$ to leave a solution to the full
equation. 

In contrast, the term with logarithmic dependence on $\hat\kappa$ can not be 
removed in this way. Hence no true nontrivial fixed point exists in this case. 
Nonetheless it is possible to cancel the logarithmic term, leaving a potential
which satisfies the RG equation and obeys the correct boundary conditions.
This can be done with the aid of the function
\begin{equation}
\hat\kappa\Bigl(\ln\hat\kappa+\ln\Lambda/\mu\Bigr),
\end{equation}
which satisfies the homogeneous version of the full RG equation (\ref{eq:dwrge3}),
\begin{equation}\label{eq:hdwrge}
\Lambda\frac{\partial}{\partial\Lambda}\left(\frac{1}{\hat V_S}\right)
=\hat p\frac{\partial}{\partial\hat p}\left(\frac{1}{\hat V_S}\right)
+\hat\kappa\frac{\partial}{\partial\hat\kappa}\left(\frac{1}{\hat V_S}\right)
-\frac{1}{\hat V_S},
\end{equation}
and so can be used to cancel the term with the logarithmic dependence on 
$\hat\kappa$. 

The resulting potential is given by
\begin{equation}\label{eq:qfp1}
\frac{1}{\hat V_0(\hat p,\hat\kappa, \Lambda)}=\hat J_0(\hat p,\hat\kappa)
+\pi\hat\kappa\Bigl({\rm Re}[H(\hat\kappa/\hat p)]+\ln\hat\kappa\Lambda/\mu\Bigr).
\end{equation}
It satisfies the DW RG equation (\ref{eq:dwrge3}), is analytic in both $\hat p^2$ 
and $\hat\kappa$, but is not a fixed point since it does depend on $\Lambda$,
albeit only logarithmically. This should not be too surprising since the general 
analysis in Sec.~{\ref{ssec:dwfp} showed that we would get a potential with a 
marginal perturbation. Although such perturbations have no power-law dependence on
$\Lambda$, in general they give rise to logarithmic evolution with $\Lambda$. If
we invert Eq.~(\ref{eq:qfp1}) we see that the potential we have obtained resums the 
leading logarithms ($\kappa\ln\Lambda$) to all orders.

Although this potential does evolve slowly with $\Lambda$, we can still expand 
around it and use the DW RG equation (\ref{eq:dwrge3}) to determine the forms of 
the perturbations around it. These are the same as the ones given above in 
Eq.~(\ref{eq:ntpert}) and so the full short-range potential is given by 
\begin{equation}\label{eq:csrp}
\frac{1}{\hat V_S}=\frac{1}{\hat V_0}+\sum_{n,m=0}^{\infty} C_{nm}
\Lambda^{m+2n-1}\hat\kappa^m \hat p^{2n}.
\end{equation}
This provides the power-counting for a strong short-range potential in the presence 
of the Coulomb potential. Note that for the $\hat\kappa$-independent terms this
is just KSW counting, which corresponds to the ERE. 

Another consequence of the marginal perturbation is the appearance in the potential 
(\ref{eq:qfp1}) of an arbitrary scale $\mu$. This perturbation is independent 
of $\Lambda$ and so cannot be separated unambiguously from a fixed-point potential. 
Its coefficient $C_{01}$ can be chosen to depend on $\mu$ in such a way that the term
\begin{equation}\label{eq:orderk}
\Bigl(\pi\ln{\hat\kappa\Lambda/\mu}+C_{01}(\mu)\Bigr)\hat\kappa
\end{equation}
and hence the full potential (\ref{eq:csrp}) are independent of $\mu$. 
(An alternative prescription would be to set $C_{01}=0$ and to take $\mu$ as the 
corresponding parameter in the potential.)

To illustrate this RG flow, we plot the behaviour of the leading coefficients in the 
expansion:
\begin{equation}\label{eq:ekexp}
\hat{V}(\hat{p},\hat{\kappa},\Lambda)=b_{00}(\Lambda)+b_{01}(\Lambda)\hat{\kappa}
+b_{20}(\Lambda)\hat{p}^2+b_{21}(\Lambda)\hat{p}^2\hat\kappa+\cdots.
\end{equation} 
In the $(b_{00},b_{20})$ plane the flow is identical to that in Fig.~\ref{fig:SRFlow} 
above. It may be interpreted in a similar way to the pure short-range case. However
the point $(-1,-1)$ is not a fixed point in the $b_{01}$ direction. This
can be seen from the flow in the $(b_{00},b_{01})$ plane shown in 
Fig.~\ref{fig:CFlow}. The only fixed point in this plane is the trivial one.

As above, the bold lines show the flows as $\Lambda\rightarrow 0$ corresponding to 
the RG eigenvectors. The flow associated with the marginal perturbation
carries the potential up the vertical flow line $b_{00}=-1$ at a logarithmic rate. 
Although the coefficient $b_{01}$ appears to be tending to infinity along this flow 
line, it will eventually become so large that the expansion in Eq.~(\ref{eq:ekexp})
breaks down. A general potential, for which $b_{00}$ is not exactly $-1$, will
flow either into the trivial point or to infinity. However, provided the coefficient
of the unstable perturbation is small, it is still possible to expand around the
logarithmic flow line, at least until some critical value of $\Lambda$.

\begin{figure}
\includegraphics*[80,510][420,720]{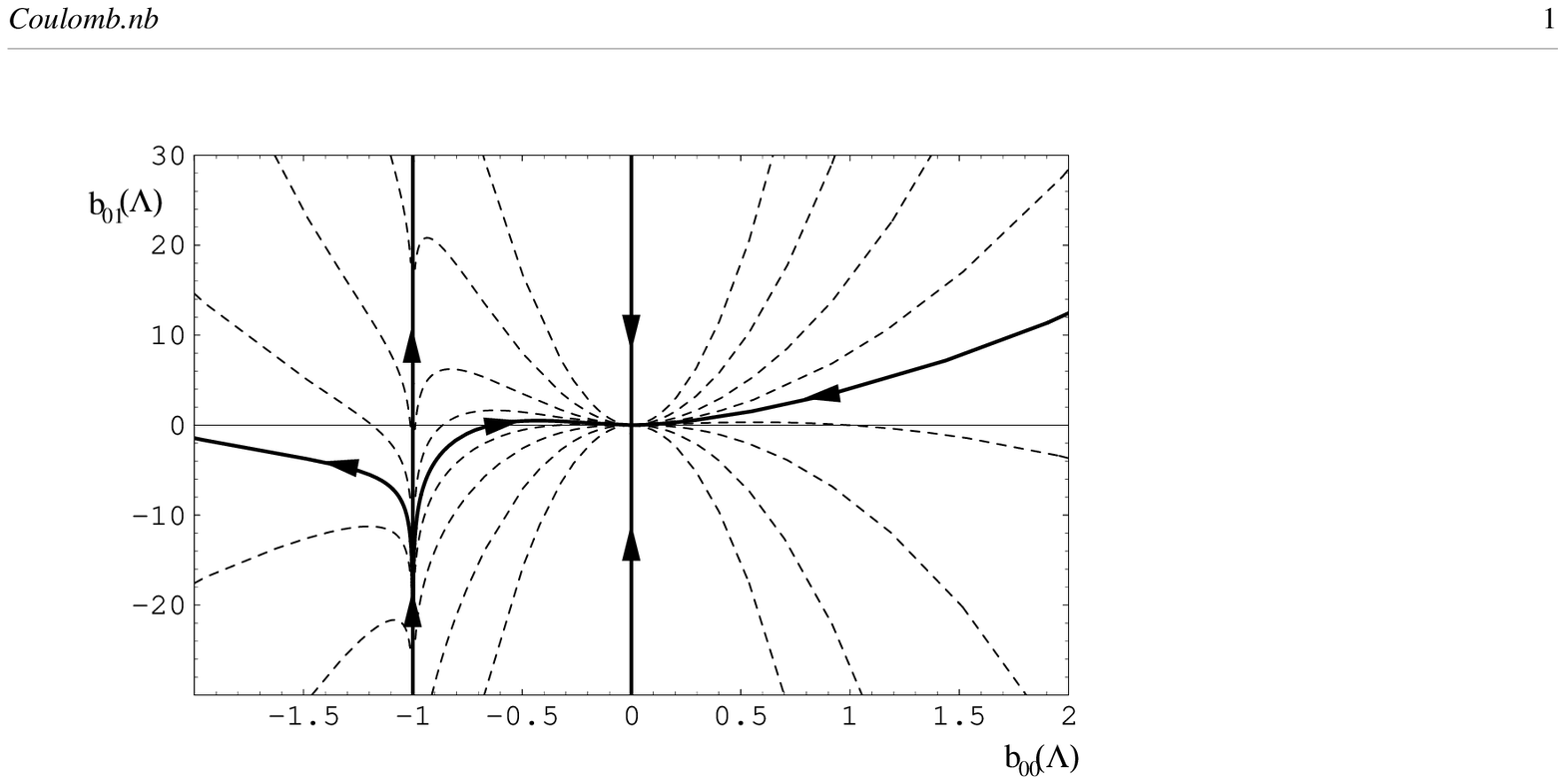}
\caption{The flow as $\Lambda\rightarrow 0$ of the coefficients 
$b_{00}$, $b_{01}$ in the expansion Eq.~(\ref{eq:ekexp}) of the
short-range potential.}\label{fig:CFlow}
\end{figure}

Substituting the effective potential into equation (\ref{eq:dwexp1}), we get the 
following expansion for DW scattering observables:
\begin{equation}
{\cal C}(\kappa/p)p\cot{\bar\delta_{S}}=-2\kappa\Bigl({\rm Re}[H(\kappa/p)]
+\ln\kappa/\mu\Bigr)+\frac{2}{\pi}\sum_{n,m=0}^{\infty}C_{nm}\kappa^{m}p^{2n},
\end{equation}
where $\kappa=\alpha M/2$. This result can now be compared with the DWERE
for the Coulomb potential \cite{bj,bethe,kr},
\begin{equation}\label{eq:coulere}
{\cal C}(\kappa/p)p(\cot{\tilde\delta_{S}}-i)+\alpha MH(\kappa/p)=-\frac{1}{\tilde a}
+\frac{1}{2}\tilde r_{e}p^{2}+\cdots,
\end{equation}
where the imaginary parts of the two terms on the left-hand side of this equation 
cancel exactly. Apart from the logarithmic term, the nonanalytic effects of the 
long-range Coulomb potential are contained in the functions ${\cal C}(\kappa/p)$ and 
$H(\kappa/p)$.

As in the pure short-range case, there is a direct correspondance between the 
expansion of the potential in powers of the energy and the terms in the 
ERE. The full effective potential gives a further expansion of 
this in powers of $\kappa$, or equivalently $\alpha$. This is organised according to 
the power counting around the logarithmically evolving potential of 
Eq.~(\ref{eq:qfp1}).
Equating the two expansions, we get the following expressions for the Coulomb-modified
scattering length and effective range,
\begin{eqnarray}
\frac{1}{\tilde a}&=&\alpha M\ln\left(\frac{\alpha M}{2\mu}\right)
-\frac{2}{\pi}\sum_{m=0}^{\infty}C_{0m}\left(\frac{\alpha M}{2}\right)^m,\\
\tilde r_e&=&\frac{4}{\pi}\sum_{m=0}^{\infty}C_{1m}\left(
\frac{\alpha M}{2}\right)^{m},
\end{eqnarray}
We see that for $\alpha=0$ these reduce to the forms given in Eq.~(\ref{eq:erecfs}) 
for the coefficients in the ordinary ERE.

The scattering length defined in Eq.~(\ref{eq:coulere}) is independent of $\mu$,
provided $C_{01}(\mu)$ is chosen as discussed above (Eq.~(\ref{eq:orderk})). However,
it still contains a logarithmic dependence on $\alpha$ and so cannot be a pure 
short-range effect. One could define a strong scattering length by
\begin{equation}\label{eq:arel}
\frac{1}{\bar a}=\frac{1}{\tilde a}-\alpha M\ln\left(\frac{\alpha M}{2\mu}\right)
=-\frac{2}{\pi}\left(C_{00}+C_{01}(\mu)\frac{\alpha M}{2}+\cdots\right),
\end{equation}
but this would depend on the arbitrary scale $\mu$. This expression may be compared 
with that of Jackson and Blatt \cite{jb},
\begin{equation}\label{eq:BJ}
\frac{1}{a_{pn}}=\frac{1}{a_{pp}}-\alpha M\Bigl(\ln{\alpha Mr_{0}}+0.33\Bigr),
\end{equation}
relating the proton-proton DW scattering length and the neutron-proton scattering 
length. Their relation assumes that the strong interaction is charge independent 
and so one might naively interpret any deviation from it as a signal of isospin 
violation. However from Eq.~(\ref{eq:arel}), we see that Jackson and Blatt's form 
corresponds to a particular choice of the scale $\mu$. A different choice of $\mu$
would lead to a different estimate of the charge-independence breaking and so
Eq.~(\ref{eq:BJ}) must be treated with caution.

\subsection{Yukawa potential}\label{sec:yukawa}

The renormalisation of scattering in the presence of a Yukawa potential is very 
similar to the Coulomb case since they both have a $1/r$ singularity at the origin 
and so their distorted waves have the same short-distance behaviour.
We consider in particular the potential for one-pion exchange (OPE) between nucleons 
in the $^1S_0$ channel,
\begin{equation}
V_L(r)=-\alpha_\pi\frac{e^{-m_\pi r}}{r},
\end{equation}
where the strength of the potential is
\begin{equation}\label{eq:alpi}
\alpha_\pi=\frac{g_A^2 m_\pi^2}{16\pi f_\pi^2}
\end{equation}
in terms of the pion mass, $m_\pi=140$ MeV, the pion decay constant, $f_\pi=93$ MeV, 
and the axial coupling of the nucleon, $g_A=1.26$.

In this case, the scaling behaviour depends crucially on the quantities we choose 
to identify as low-energy scales. As in ChPT, the pion mass clearly provides one 
such scale. Treating this scale explicitly will allow us to make a chiral expansion
of an EFT which we might hope would be valid for momenta below the $\rho$-meson 
mass.

When rescaled the OPE potential is, in momentum space,
\begin{equation}
\hat V_L(\hat{\bf k}',\hat{\bf k},\hat m_\pi,\Lambda)
=-\Lambda\frac{Mg_A^2}{8\pi^2f_\pi^2}
\frac{\hat m_\pi^2}{|\hat{\bf k}'-{\bf k}|^2+\hat m_\pi^2}.
\end{equation}
This is proportional to the cut-off $\Lambda$ and so its effect on the RG flow 
vanishes as $\lambda\rightarrow 0$. This means that the fixed points will be the
same as in the pure short-range case. The DW Green's function $G_L$ can be expanded
in powers of the long-ranged potential and pion exchange treated as perturbative
corrections to the ordinary ERE. This is the KSW scheme for including pions explicitly
in an EFT for nucleon-nucleon scattering \cite{ksw2,ksw3}.

Although the KSW scheme allows ChPT to be extended to the two-nucleon sector, the
resulting expansion turns out to be, at best, slowly convergent 
\cite{sf,ch,krthesis,fms}. The problem is that the pion-nucleon coupling is large,
or equivalently that the scale which sets the the strength of the potential,
\begin{equation}
\lambda_{NN}=\frac{16\pi f_\pi^2}{M g_A^2}\simeq 300\ \rm{MeV},
\end{equation}
is small. There is thus no good separation of scales and so it is
unsurprising that the corresponding EFT shows poor convergence.

An alternative approach is to treat $\lambda_{NN}$ or equivalently the inverse 
``pionic Bohr radius",
\begin{equation}
\kappa_\pi=\frac{\alpha_\pi M}{2},
\end{equation}
as an additional low-energy scale. The rescaled OPE potential is then independent
of $\Lambda$ and so has been promoted to form part of any fixed point. 

The effects of the distortion show up in the normalisation of the DW's near the
origin. Although it is not possible to write this normalisation in analytic form, 
from dimensional analysis, it must be of the form
\begin{equation}
|\psi_L(p,R\rightarrow 0)|^2={\mathcal{C}}_\pi(\kappa_\pi/p, m_\pi/p).
\end{equation}
If we rescale both $\kappa_\pi$ and $m_\pi$, as just discussed, the normalisation 
of the DW with $p=\Lambda$ is independent of $\Lambda$. The resulting RG equation is 
\begin{equation}
\Lambda \frac{\partial \hat V_S}{\partial\Lambda}=
\hat p \frac{\partial \hat V_S}{\partial \hat p}+
\hat \kappa_\pi \frac{\partial \hat V_S}{\partial \hat \kappa_\pi}+
\hat m_\pi \frac{\partial \hat V_S}{\partial \hat m_\pi}+
\hat V_S+\frac{{\mathcal{C}}_{\pi}(\hat \kappa_\pi,\hat m_\pi)}{1-\hat p^2}
\hat V_S^2.
\end{equation}
The expansion around the trivial fixed point is
\begin{equation}
\hat V_S=\sum_{l,m,n}C_{lmn}\Lambda^\nu 
\hat m_\pi^{2l} \hat \kappa_\pi^m \hat p^{2n},
\end{equation}
where the eigenvalues are $\nu=2l+m+2n+1=1,3,5,\cdots$, as in the Coulomb case.

The more interesting expansion around a nontrivial fixed point takes the form
\begin{equation}
\frac{1}{\hat V_S}=\frac{1}{\hat V_0}+\sum_{l,m,n} C_{klm}\Lambda^\nu 
\hat m_\pi^{2l} \hat \kappa_\pi^m \hat p^{2n},
\end{equation}
where $\nu=2l+m+2n-1=-1,0,+1,\cdots$. This includes a marginal perturbation 
which is linear in the inverse Bohr radius and leads to logarithmic evolution of
the potential $\hat V_0$ with $\Lambda$. We can resum this using the RG method 
described above in Sec.~\ref{sec:coulomb}. The presence of one unstable 
perturbation, as well as the marginal one, means that the RG flows are very similar 
to those shown in Figs.~\ref{fig:SRFlow} and \ref{fig:CFlow}.

The expansion of the potential corresponds to a DWERE of the form
\begin{equation}
\mathcal{C}_\pi(\kappa_\pi/p,m_\pi/p)\cot{\delta_S}
=2\kappa_\pi\left(H_\pi(\kappa_\pi/p,m_\pi/p)
+\ln{\kappa_\pi/\mu}\right)+\frac{2}{\pi}\sum_{l,m,n}
C_{lmn}m_\pi^{2l}\kappa_\pi^m p^{2n},
\end{equation}
where the dependence on the arbitrary scale cancels between the first term
and $C_{010}(\mu)$. All nonanalytic behaviour is contained in the functions
$\mathcal{C}_\pi(\kappa_\pi/p,m_\pi/p)$ and $H_\pi(\kappa_\pi/p,m_\pi/p)$.
For the Yukawa potential these must be calculated using numerical methods.
This DWERE has been applied to nucleon-nucleon scattering in 
Refs.~\cite{sf,krthesis}.

The EFT corresponding to this DWERE is one in which the effects of OPE and an 
energy-independent short-range force are iterated to all orders. The
resummation of these two terms, which would be of leading order in Weinberg
power counting about a trivial fixed point, is precisely the scheme
suggested by Weinberg in his original paper \cite{wein2} on EFT's for 
nucleon-nucleon scattering. This WvK scheme was then developed and applied by 
van Kolck and others \cite{orvk,vk1,vk4,egm}, with some success. The RG 
analysis shows that the expansion around the logarithmically evolving potential 
defines the power counting for this scheme, a term $m_\pi^{2l}\alpha_\pi^m p^{2n}$ 
being of order $d=2l+m+2n-2$. For the terms which depend on energy only, this 
counting is similar to that of the KSW scheme.

Although the original motivation for developing EFT's for nucleon-nucleon scattering 
was to extend ChPT to few-nucleon systems, it is not obvious that the WvK scheme is 
consistent with the chiral expansion. For example the term proportional to 
$\alpha_\pi$ is of order $d=-1$ in this scheme, whereas a term proportional to
$m_\pi^2$ is of order $d=0$. Since $\alpha_\pi$ is proportional to $m_\pi^2$, both
terms would be of the same order in a pure chiral expansion. Within the chiral
expansion the quantity $\lambda_{NN}$ would be a high-energy scale, but in the 
WvK scheme it has been treated as a low-energy scale. Hence in contrast to the
the KSW scheme, the direct link to ChPT has been lost.

\subsection{Repulsive inverse-square potential}\label{sec:repisq}

As a final example we consider the inverse-square potential, which is of interest 
because of its relevance to the three-body problem. Efimov \cite{efim} has shown 
that this potential can describe the scattering of a particle off a two-particle 
bound state in the limit where the two-body potential has zero range and infinite 
scattering length. The renormalisation of the short-ranged three-body forces
which appear in EFT treatments of such systems is currently the object of
keen study \cite{bhvk,bhk}.

In coordinate space the potential has the form 
\begin{equation}
V_L(r)=\frac{\beta}{M r^2}.
\end{equation}
When an overall factor of $1/M$ is taken out, as here, we see that the remaining
coupling $\beta$ is dimensionless. The potential is thus scale-free and so it 
can natuarally be thought of as part of a fixed point. This is why it appears in 
the three-body problem with a scale-free two-body interaction.

The inverse-square potential acts like the centrifugal term in the free
Schr\"odinger equation and so the results we obtain here can also be applied to 
scattering in higher partial waves by a short-range potential. In general, the 
distorted waves are given in terms of Bessel functions of noninteger order and
have the form
\begin{equation}\label{eq:invsqdw}
\psi_L(p,r)=A_1r^{-1/2}J_{\nu}(pr)+A_2r^{-1/2}J_{-\nu}(pr),
\end{equation}
where the order $\nu=\sqrt{1/4+\beta}$ depends on the strength of the coupling. 
This makes it clear that two cases need to be considered separately, real $\nu$
($\beta>1/4$) and imaginary $\nu$ ($\beta<1/4$). In this paper we consider only
the repulsive inverse-square potential. The more complicated case of a strongly 
attractive attractive potential, where $\nu$ is imaginary, will be discussed in a 
future work. 

In the repulsive case, the relevant solution, which is regular at the origin and 
correctly normalised, is
\begin{equation}
\psi_L(p,r)=\sqrt{\frac{\pi}{2pr}}J_{\nu}(pr).
\end{equation}
The corresponding $s$-wave phase-shift is easily determined to be 
$\delta_L=\pi/4-\pi\nu/2$. To construct the RG, we need the
square of the DW at small $R$,
\begin{equation}
|\psi_L(p,R)|^2=\frac{1}{\Gamma(1+\nu)^2}\left(\frac{p R}{2}
\right)^{2\nu-1}.
\end{equation}
This is of the form in Eq.~(\ref{eq:wfpl}) with $\sigma=2\nu$ and
${\cal N}^2=1/[2^{2\nu-1}\Gamma(1+\nu)^2]$. For scattering in a partial wave with
$l>0$, the centrifugal barrier provides a $1/r^2$ potential with $\beta=l(l+1)$
and hence $\nu=l+\frac{1}{2}$.

Since ${\cal N}^2$ is simply a constant in this case, it is convenient to absorb it 
into the rescaling of the potential. The resulting RG equation is then 
\begin{equation}\label{eq:isrge}
\Lambda\frac{\partial\hat V_S}{\partial\Lambda}
=\hat p\frac{\partial\hat V_S}{\partial\hat p}
+2\nu\hat V_S+\frac{1}{1-\hat p^2}
\hat V^2_S.
\end{equation}
From the general analysis in Sec.~\ref{sec:rglong}, we see that perturbations
around the trivial fixed are of the form, 
\begin{equation}\hat{V}_S=C_{2n}\Lambda^{2n+2\nu}p^{2n}. 
\end{equation}
Since $\nu>0$ all of the eigenvalues are positive and the fixed point is stable as 
$\Lambda\rightarrow 0$. The term proportional to $p^{2n}$ is of order $d=2n+2\nu-1$ 
in the corresponding power counting. For nucleon-nucleon scattering in partial waves
with $l>0$ there are no bound states or resonances close to threshold and so this
fixed point is the appropriate one. The power counting for this case is given by
$d=2(n+l)$.

Also of interest is the nontrivial fixed point, which corresponds to a DWERE.
To construct it, we start from the basic loop integral of Eq.~(\ref{eq:bloop}). 
The cases of integer and noninteger $\nu$ behave differently and need to be 
considered separately. The loop integral can be evaluated to give
\begin{equation}
\hat J_0(\hat p)={\cal P}\int_0^1 \hat q^{2\nu+1} d\hat q \frac{1}{\hat p^2-
\hat q^2}=\frac{1}{2}\sideset{}{'}\sum_{n=0}^\infty\frac{\hat p^{2n}}{n-\nu}
+\frac{\pi}{2}\hat p^{2\nu}G(\hat p,\nu),
\end{equation}
where,
\begin{equation}
G(\hat p,\nu)=
\begin{cases}
\cot{\pi\nu}, &\nu\not\in{\mathbb{N}}\\
2\ln\hat p, & \nu\in{\mathbb{N}}.
\end{cases} 
\end{equation}
The prime on the sum here indicates that the term with $n=\nu$ term must be omitted 
when $\nu$ is an integer.

To obtain a well-behaved short-range potential, we apply the boundary condition 
of analyticity by cancelling the final, nonanalytic term from $\hat J_0$. When $\nu$
is noninteger we can do this using the fact that $p^{2\nu}$ satisfies the 
homogeneous version of the RG equation (\ref{eq:isrge}). In the case of integer 
$\nu$, the logarithm of $\hat p$ can be cancelled in similar manner to the logarithm 
of $\hat\kappa$ which appears for the Coulomb potential. This leads to a potential 
with a logarithmic dependence on $\Lambda$, in which the leading logarithms are 
resummed to all orders. The result in either case can be expressed in the form
\begin{equation}
\frac{1}{\hat V_0(\hat p)}=\hat J_0(\hat p)-\frac{1}{2}\hat p^{2\nu}
G(\hat p \Lambda/\mu,\nu).
\end{equation}           
Once again the full solution to the RG equation is obtained by adding perturbations 
around this fixed point, 
\begin{equation}
\frac{1}{\hat V_S}=\frac{1}{\hat V_0}+\sum_{n=0}^{\infty}C_{2n}
\Lambda^{2n-2\nu}\hat p^{2n}.
\end{equation}
This fixed point is unstable, with the number of negative eigenvalues being 
governed by $\nu$. If $\nu$ lies between the integers $N-1$ and $N$, the
first $N$ perturbations are unstable. If $\nu=N$ then there is also a marginal 
eigenvector, $\hat p^{2N}$, which is the origin of the logarithmic behaviour.
The corresponding coefficient $C_{2N}(\mu)$ depends on the arbitrary scale $\mu$
so that the full potential is $\mu$-independent.

Figs.~3 and 4 show this flow for the cases $\nu=0.85$ and $\nu=1.15$ respectively.
We again expand the potential in powers of energy, as in Eq.(\ref{eq:enexp}), and
plot the RG flow in the $(b_0,b_2)$ plane. For $\nu=0.85$ there is only one unstable 
perturbation. However the flow in the direction of the lowest stable perturbation
is quite weak, $\propto\Lambda^{0.3}$, and so the flow lines peel away from the 
critical surface much more rapidly than in Fig.~1. This flow weakens as 
$\nu\rightarrow 1$ until at $\nu=1$ it becomes the logarithmic flow of a marginal
perturbation. For $\nu>1$, the lowest two perturbations are unstables, as shown
in Fig.~4.

\begin{figure}
\includegraphics*[70,560][420,775]{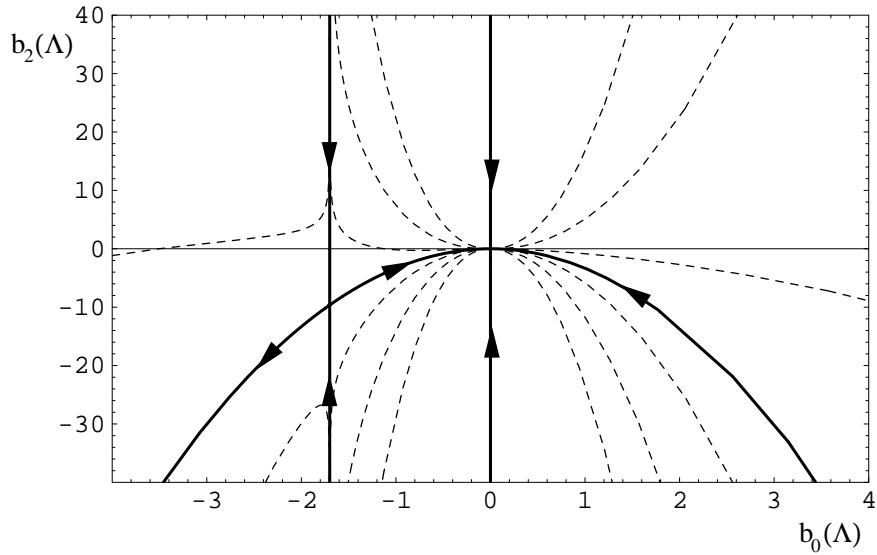}
\caption{The flow as $\Lambda\rightarrow 0$ of the first two coefficients in the 
expansion Eq.~(\ref{eq:enexp}) of the short-range potential in the presence of
an inverse-square potential with $\nu=0.85$.}
\label{InvSquFlow1}
\end{figure}

\begin{figure}
\includegraphics*[70,560][420,775]{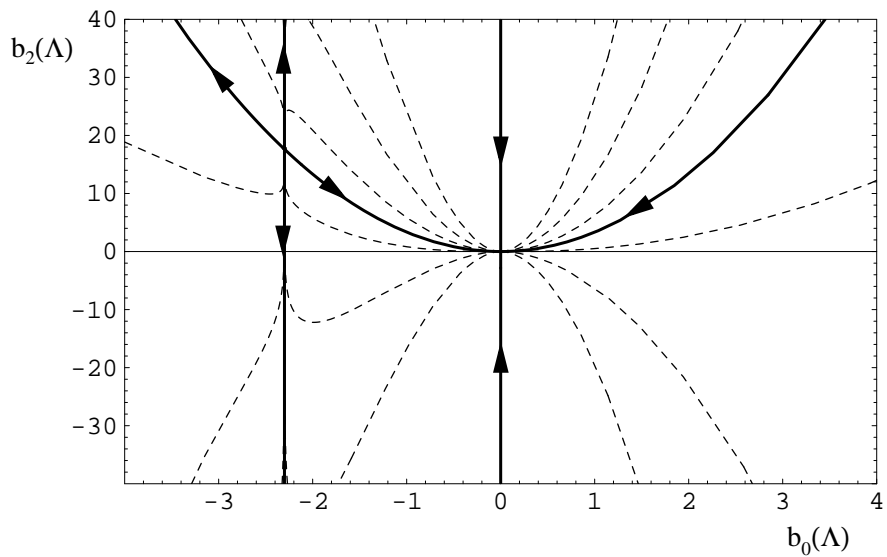}
\caption{As Fig.~3 but for $\nu=1.15$}
\label{InvSquFlow2}
\end{figure}

The power counting around the nontrivial fixed point is $d=2n-2\nu-1$ for a term
proportional to $p^{2n}$. This is quite different from the counting found for
scattering in the presence of the Coulomb potential. Since the inverse-square 
potential is scale-free, its strength does not provide an expansion parameter in the 
low-energy EFT. Instead it appears in the energy power-counting itself, determining 
the number of relevant (unstable) perturbations. In the limit where this strength
vanishes, and $\nu\rightarrow \frac{1}{2}$, we have precisely the power-counting 
established earlier for a pure short-range potential.

The scattering amplitude for this potential can be calculated as in 
Sec.~\ref{sec:dwere}. The result can then be expanded the form of a DWERE as
\begin{equation}
p^{2\nu}\Bigl(\cot{\delta_S}-\frac{1}{\pi}G(p/\mu,\nu)\Bigr)
=\frac{2}{\pi}\sum_{n=0}^{\infty}C_{2n}p^{2n}.
\end{equation}
This is an expansion in powers of the energy, which is the only scale in this 
system. In general the coefficients have unusual, noninteger dimensions, as a result 
of the noninteger power of the energy on the left-hand side. For example the 
leading coefficient, which is the analogue of a modified scattering length, has a 
dimension of $2\nu$.

In the case of scattering of a particle with angular momentum $l$ by a 
short-range potential, we have $\nu=l+\frac{1}{2}$ and there is no nonanalytic 
energy dependence in $\cot\delta_S=\cot(\delta+l\pi/2)$. The ERE becomes 
\begin{equation}
p^{2l+1}\cot\left(\delta+\frac{l\pi}{2}\right)
=\frac{2}{\pi}\sum_{n=0}^{\infty}C_{2n}p^{2n}.
\end{equation}
For $l=0$ this is the familiar $s$-wave ERE given above in Eq.~(\ref{eq:ere}).
For $l=1$ we get the $p$-wave expansion, whose leading term is a scattering
volume rather than a length.

\section{Discussion}

Over the last few years, EFT's have been developed for low-energy scattering of two 
heavy particles by short-range interactions. In particular these have been 
successfully applied to nucleon-nucleon scattering. The techniques described here 
extend the RG ideas which underly these theories to systems where the particles 
interact by a combination of known long-range and unknown short-range forces.
These provide a framework for constructing EFT's for such systems.

By expanding the short-range potential about a fixed point of the RG, we can 
systematically organise the terms in the potential according a power counting
defined by their RG eigenvalues. These eigenvalues also determine whether a 
perturbation is stable, unstable or marginal.

A crucial feature of our approach is that we regulate the loop integrals by
cutting them off in the basis of DW's for the long-range potential. This ensures 
that the long-range potential is not modified by the cut-off. As a result the RG 
equation has a simple form, which is very similar to that found for a pure 
short-range potential.

Another important feature is the need to identify all low-energy scales associated
with the long-range potential. Our method can be applied when the resulting rescaled
potential is independent of the cut-off and so it can be treated as part of a fixed
point, its effects being resummed to all orders in the DW's. If the rescaled 
long-range potential diverges as the cut-off is lowered, then no fixed point 
can be found. On the other hand, if the rescaled potential vanishes in this limit,
then it can be treated as a perturbation about a fixed point of the RG for a pure 
short-range potential.

As in the case of a pure short-range potential, we always find a trivial fixed point.
The expansion around this point can be used to describe systems where the additional 
scattering by the short-range potential forms a small correction to the effect of
the long-range potential. The terms in this potential corespond to an expansion
of the DW $K$-matrix in powers of the energy and any other low-energy scales.

In some cases we also find an energy-dependent potential which forms a nontrivial 
fixed point of the RG. In other cases, such a fixed point would have a marginal
perturbation and instead we find a potential which evolves logarithmically with
the cut-off. These potentials describe systems which have bound states lying exactly
at threshold and the expansions around them correspond to DW versions of the 
effective-range expansion.

We have applied this method to several examples. An effective field theory for the 
strong interaction in proton-proton scattering can be constructed by considering 
the Coulomb potential as the long-range distorting potential. The inverse Bohr 
radius then provides an additional low-energy scale. This is a case where a 
nontrivial fixed point would have a marginal perturbation and the RG can be used 
to resume the resulting logarithmic corrections. The expansion around this potential
corresponds to the DWERE for the Coulomb potential, where the coefficents in the
expansion in powers of the energy have themselves been expanded in powers of the
inverse Bohr radius. The logarithmic evolution of the potential shows up in the
fact that the purely short-range scattering is not uniquely defined, as it depends
logarithmically on an arbitrary scale.

The RG analysis of the scattering in the presence of a Yukama potential is similar
to that for the Coulomb one. In the specific example of nucleon-nucleon scattering
in the $^1S_0$ channel, the Yukawa potential is provided by OPE. Here we clearly 
want to identify the pion mass as a low-energy scale if we wish to make contact with 
ChPT. However we have a choice about how we treat the scale
\begin{equation}
\kappa_\pi=\frac{g_A^2 M m_\pi^2}{32\pi f_\pi^2}.
\end{equation}
In strict chiral counting this inverse ``Bohr radius" is of second order in $m_\pi$.
Treating it in this way, the rescaled OPE potential vanishes linearly with the
cut-off and so can be treated a perturbation in an EFT based on a fixed point
of a pure short-range potential. This is is the basis for the KSW scheme for 
including pion-exchange forces in EFT's for nucleon-nucleon scattering.

The alternative is to regard $\kappa_\pi$ as a low-energy scale, analogous to
the treatment of the inverse Bohr radius in the Coulomb case. The rescaled Yukawa
potential is then independent of cut-off and should be resummed. This is the WvK 
scheme for including OPE. The EFT based on the resulting nontrivial (logarithmically
evolving) potential is equivalent to a DWERE. However, although this scheme results
in a consistent EFT, the treatment of $\kappa_\pi$ as a quantity of first-order in 
low-energy scales means that the connection with the chiral expansion has been lost.

Our final example is the inverse-square potential which is of particular interest 
becuase of its relevance to three-body scattering. We have considered here the 
case of repulsive potentials, which are relevant to three-body systems such as
neutron-deuteron scattering with $J=\frac{3}{2}$, and also to two-body scattering in 
higher partial waves. This potential is scale free and so its strength shows up in 
the RG eigenvalues, and hence the power counting. Although it is possible to find 
a nontrivial fixed point corresponding to a DWERE, the number of unstable 
perturbations increases with the strength of the long-range potential. This implies 
that an extremely delicate fine-tuning would be required to generate a bound state 
at threshold. In general one would expect scattering in such systems to be weak, as
it is for neutron-proton scattering in high partial waves. The appropriate EFT's 
for them are based on the (stable) trivial fixed point.

We are currently extending this approach to describe scattering in the presence of
an attractive inverse-square potential, which is more complicated than the repulsive 
case since the DW's oscillate rapidly near the origin. The resulting EFT's 
will be relevant to three-body systems such as neutron-deuteron scattering with 
$J=\frac{1}{2}$. It will also be interesting to explore more singular potentials,
such as OPE in the $^3S_1$ channel. Finally the real power of the EFT's is their
ability to form direct connections between observables for different processes. Doing
this will require enlarging the present treatment to included couplings to external 
electromagnetic and weak currents.

\section*{Acknowledgments}

We wish to thank J. McGovern for helpful discussions and a critical reading of 
this paper. MCB is grateful to the Institute for Nuclear Theory, Seattle for 
hospitality during the programme on Effective Field Theories and Effective 
Interactions where these ideas first took shape, to S. Beane, T. Cohen, S. Coon,
H. Griesshammer, D. Phillips, U. van Kolck and others for useful
discussions, and to K. Richardson for an introduction to DW effective-range
expansions. This work was supported by the EPSRC.

\end{document}